\documentclass{article}
\PassOptionsToPackage{numbers,sort&compress}{natbib} %j edit
\usepackage[preprint]{neurips_2025}
\usepackage[utf8]{inputenc} % allow utf-8 input
\usepackage[T1]{fontenc}    % use 8-bit T1 fonts
\usepackage{hyperref}       % hyperlinks
\usepackage{url}            % simple URL typesetting
\usepackage{booktabs}       % professional-quality tables
\usepackage{amsfonts}       % blackboard math symbols
\usepackage{nicefrac}       % compact symbols for 1/2, etc.
\usepackage{microtype}      % microtypography
\usepackage{xcolor}         % colors

\usepackage{graphicx}
\usepackage{amsmath}
\usepackage{float}
\usepackage{subcaption}
\usepackage{multirow}
\usepackage{tabularx}
\usepackage{caption}

\newif\iflatexml
\ifdefined\LaTeXML        % LaTeXML sets \LaTeXML
  \latexmltrue
\else
  \latexmlfalse
\fi
\title{NLP4Neuro: Sequence-to-sequence learning for neural population decoding}

\author{%
  \textbf{Jacob J.~Morra}$^{\dagger}$\quad
  \textbf{Kaitlyn E.~Fouke}\quad
  \textbf{Kexin Hang}\quad
  \textbf{Zichen He}\quad
  \textbf{Owen Traubert}\\[0.1em]
  \textbf{Timothy W.~Dunn}\quad
  \textbf{Eva A.~Naumann}\\[0.4em]
  Duke University, Durham NC 27705, USA\\
  $^{\dagger}$\texttt{jacob.morra@duke.edu}
}

\begin{document}

\maketitle

\begin{abstract}
Delineating how animal behavior arises from neural activity is a foundational goal of neuroscience. However, as the computations underlying behavior unfold in networks of thousands of individual neurons across the entire brain, this presents challenges for investigating neural roles and computational mechanisms in large, densely wired mammalian brains during behavior. Transformers, the backbones of modern large language models (LLMs), have become powerful tools for neural decoding from smaller neural populations. These modern LLMs have benefited from extensive pre-training, and their sequence-to-sequence learning has been shown to generalize to novel tasks and data modalities, which may also confer advantages for neural decoding from larger, brain-wide activity recordings. Here, we present a systematic evaluation of off-the-shelf LLMs to decode behavior from brain-wide populations, termed NLP4Neuro, which we used to test  LLMs on simultaneous calcium imaging and behavior recordings in larval zebrafish exposed to visual motion stimuli. Through NLP4Neuro, we found that LLMs become better at neural decoding when they use pre-trained weights learned from textual natural language data. Moreover, we found that a recent mixture-of-experts LLM, DeepSeek Coder-7b, significantly improved behavioral decoding accuracy, predicted tail movements over long timescales, and provided anatomically consistent highly interpretable readouts of neuron salience. NLP4Neuro demonstrates that LLMs are highly capable of informing brain-wide neural circuit dissection.
\end{abstract}

\section{Introduction}
Predicting behavior from neuron-scale network activity, or neural decoding, remains an open challenge in computational neuroscience; one which promises to improve our understanding of the function of neural circuitry in various model organisms, characterize individual differences, and advance our capacity to build brain-computer interfaces for treating neural dysfunctions affecting brain-wide circuits, such as in various neurodevelopmental or neurodegenerative disorders \cite{Tayebi2023-ed, Mathis2024-fm}. 

Machine learning (ML) is an increasingly used tool for neural decoding tasks, for two reasons: First, ML can be applied to myriad domains without the level of expertise required to build models from scratch; second, it achieves high fidelity performance across task types \cite{Mathis2024-fm}. In other words, it works well as a general-purpose tool and requires minimal user interference. 

Taking these benefits further, recent progress in ML has leveraged pre-training; that is, learning from an often very large dataset to acquire generalized knowledge, and fine-tuning in order to sharpen that knowledge towards a new task (transfer learning). This trend is particularly apparent in natural language processing (NLP), wherein it has been shown that generalization is an emergent property of scale in off-the-shelf large language models (LLMs) \cite{Rae2021-nq, Wei2022-kl, Liu2023-yc, Xiong2021-yu}.

It is an open question whether this emergence carries over to neural decoding, i.e. as opposed to using simpler \cite{Dixen2025-fl}, linear \cite{Zhang2023-uf}, or recurrent models; the latter being previously widely used in both NLP \cite{Wang2015-uu, Dominey2022-nz} and neuroscience \cite{Creamer2018-pg, Creamer2024-sh, Lappalainen2024-ip, Tschopp2018-iu, Bardozzo2024-vm, Morra2025-pf}. However, many contemporary approaches in neural decoding specifically have moved to larger models with higher parameter counts, away from recurrent neural networks (RNNs) and towards transformers, as in \cite{Ye2021-ry, Ye2023-ku, Zhang2024-je, Wang2023-ge, Chau2024-uj}. Moreover, progress in NLP is expeditious. With LLMs reporting impressive abilities to transfer their language knowledge across domains -- i.e. in generating candidate protein binders \cite{Chen2024-zf}, suggesting novel material design architectures \cite{Buehler2023-wo}, and learning implicit graph structures \cite{Zhang2025-yq} -- and the continued release of open-access, state-of-the-art models, pre-trained on billions of tokens \cite{Guo2024-dk}, there is an opportunity to develop a modular, LLM-driven pipeline with the goal of advancing neural decoding. 

We present a two step process, NLP4Neuro: In the first, model-building stage (Fig. \ref{fig:calcium_and_behavior}), we select from off-the-shelf LLMs, and evaluate their capacity to decode behavior from neural data. That is, to learn a function that maps, for a given sequence length $s \in \{5,10,15,20\}$, $s$ tokens of neural population activity to $s$ behavior outputs, measured simultaneously in a known model organism, the \textit{larval zebrafish}. Using data acquired from simultaneous two-photon calcium imaging and head-fixed tail behavior tracking, neuron-scale population activity is matched to high-frequency tail kinematics. This stage can continue to be updated and improved with the addition of state-of-the-art models. In the second, model-interpretation stage, we use the best model as a practical tool for finding features (neurons) which contribute most to observed behavior. In particular, we use salience mapping from the model weights to inform each neuron's spatial location and regional information \cite{Hou2023-ls, Denil2014-wb, Serrano2019-wk}.

We ground our approach in four experiments: In the first three experiments, we compare model paradigms, their pre-trained weights, and various token embedding strategies, to determine the resulting impacts on neural decoding. In experiment 1, we add linearly-trained and nonlinear RNNs to the model pipeline, alongside LLMs, to contrast RNNs and transformers, with varying parameter counts. In the second experiment, we assess the value of pre-trained weights for neural decoding. In the third experiment, we evaluate reasonable embedding strategies from the NLP domain (see Ssec. \ref{ssec:embeddings}). Pre-trained weights and input embeddings are fundamental LLM features \cite{Wang2022-vj}. In the fourth experiment, we assess the viability of our best model as an explainable neural decoding tool.

For RNN models, we first consider the reservoir computer (RC), which is best-in-class for modeling dynamical systems, whilst being linearly-trained and lightweight \cite{Gauthier2021-yv}. It is also used in NLP, i.e. in comprehension of narrative sequence structure \cite{Dominey2022-nz}. We also consider the long short-term memory network (LSTM), which was previously state-of-the-art in natural language inference \cite{Wang2015-uu}. For LLMs, we consider variations of the transformer which include encoder-only models, e.g. Bidirectional Encoder Representations from Transformers (BERT) -- the base uncased version -- from \cite{Devlin2018-ym}, decoder-only models, e.g. Generative Pretrained Transformer 2 (GPT-2) from \cite{Radford2019-kc}, and Mixture of Expert (MoE) models, i.e. DeepSeek Coder-7b (-c7b), from \cite{Guo2024-dk}, which performs similarly to GPT-3.5 Turbo in next token prediction, operating with 7b (7 billion) parameters instead of GPT-3.5's 175b. 

We report that the MoE transformer, DeepSeek-c7b, achieves the strongest performance across all sequence-lengths except for $s=15$, followed by BERT-base-uncased (-bu), whereas the LSTM is the best-performing RNN and achieves competitive scores for other sequence lengths. We find, furthermore, that DeepSeek-c7b outpaces the LSTM by a wider margin when assessing their performance from $s=5$ to $s=20$, i.e. for longer timescales. Evaluating the impact of pre-training on the performance of transformer models alone, we find that pre-trained weights generate lower-error predictions on neural data for the MoE and encoder-only models. Regarding the neural data encoding strategies we employed, we find that relative position encoding improves performance across transformer models, but also that DeepSeek-c7b is insensitive to all strategies. Finally, we use DeepSeek-c7b, the best-performing model, to illustrate the use of NLP4Neuro as a tool for identifying single-neurons as features which may exert a controlling effect on tail behavior through salience mapping.

\section{Related work}

\textbf{Supervised models with benchmarks for neural data prediction} \hspace{.5cm} Recently,  \cite{Lueckmann2025-ze} introduced the Zebrafish Activity Prediction Benchmark (ZAPBench) for predicting neuron-scale resolution activity from previous neural activity, using a range of linear and deep learning (DL) models. \cite{Lueckmann2025-ze} finds that models mostly outperform naive baselines, particularly for larger context windows and more forward prediction steps. For short context (4 time steps), a DL model (U-Net) outperforms linear and multi-layer perceptron (MLP) models. NLP4Neuro uses various LLMs to learn a functional mapping of multi-component tail behavior from neural activity data. It also works as a tool for identifying salient, tail-controlling neurons, to be later validated (i.e. with photostimulation). Myriad other works have used neural circuitry to constrain and predict behaviors and their dynamics \cite{Musall2019-rv, Andalman2019-dz, Haesemeyer2018-fr, Urai2022-cf, Vyas2020-wr}.

\textbf{Transformers applied to neural population decoding} \hspace{0.5cm} With POYO (Pre-training On manY neurOns) and POYO+, \cite{Zhang2024-je} and \cite{Azabou2024-cu} encode neural action potentials into neural data tokens for monkey, human, and mouse data. POYO+ is scalable across sessions tasks. Latent Factor Analysis via Dynamical Systems or LFADS, similarly to POYO and POYO+, reduces spiking neural data to a low-dimensional set of per-trial features to make predictions on synthetic and neural spike train activity data. \cite{Ye2021-ry, Ye2023-ku, UnknownUnknown-yw} present Neural Data Transformers (NDTs), the latest of which pairs macroscale neural population activity with motor covariates in monkey and human subject data. The Population Transformer (PopT), introduced in \cite{BarbuUnknown-yd}, generates ensemble-level and channel-level human brain neural population representations at macroscale, for decoding subject tasks (e.g. sentence onset) and building functional connectivity maps. Conversely, we (1) introduce a neuron-scale, brain-wide neural decoding approach and (2) explicitly use off-the-shelf, pre-trained LLMs to accomplish this. 

\section{Methods}
\subsection{Data Acquisition} 
For all experiments, we used 6-day-old, transgenic zebrafish \textit{Tg(elavl3:H2B-GCaMP6s)} expressing the green calcium indicator GCaMP6s in most neurons across the brain. Using volumetric two-photon microscopy, we measured neural activity across the mid and hindbrain in head fixed zebrafish in 2\% weight/volume low melting point agarose in embryo medium, with the tail freed for high-speed behavior tracking with an infrared-sensitive camera. We imaged 5 planes (512 x 512 pixels), 7 $\mu$m apart, spanning a 308 $\mu$m ($x$) $\times$ 308 $\mu$m ($y$) $\times$ 35 $\mu$m ($z$) = $3.3 \times 10^5 \mu\text{m}^3$ volume, sampled at 1.1 Hz. This field of view included the visually responsive, retinorecipient pretectum (Pt), spinal projection neurons residing in the nucleus of the medial longitudinal fasciculus (nMLF), and other motor command neurons in the hindbrain (Hb). During imaging (Fig. \ref{fig:expsetup}), freely swimming zebrafish were presented with a sequence of drifting stimuli evoking optomotor response behaviors \cite{Fouke2025-xy}. All experiments adhered to institutional guidelines.

% ---------- figure ----------
\iflatexml
  \begin{figure}[!tbp]
    \centering
    \begin{subfigure}[t]{0.275\textwidth}
        \centering
    \end{subfigure}
    \begin{subfigure}[t]{0.275\textwidth}
        \centering
        \raisebox{1.25mm}[0pt][0pt]{%
        \includegraphics[width=1\linewidth]{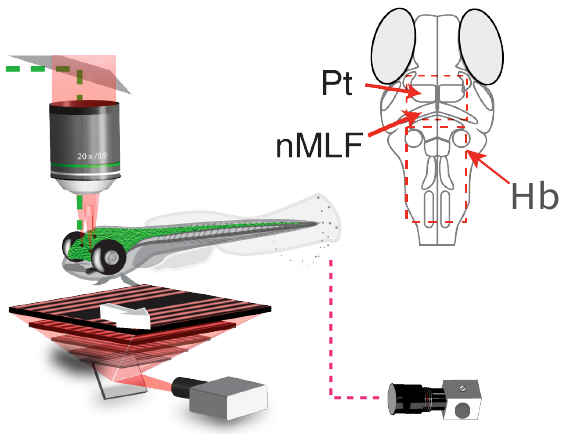}
        }
        \caption{}
        \label{fig:expsetup}
    \end{subfigure}
    \begin{subfigure}[t]{0.275\textwidth}
        \centering
    \end{subfigure}
    \\[1ex]
    \begin{subfigure}[t]{0.6\textwidth}
        \centering
        \includegraphics[width=1\linewidth]{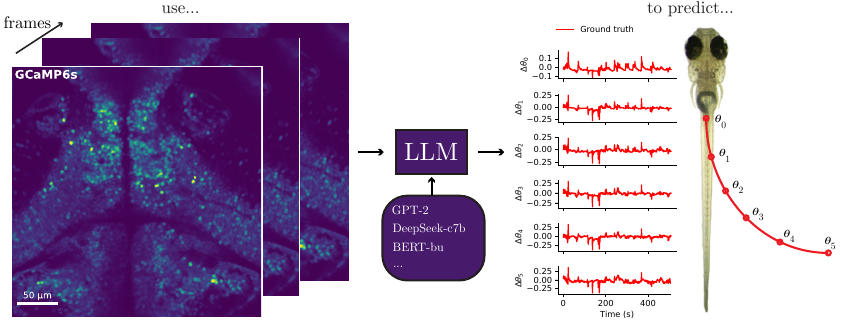}
        \caption{}
        \label{fig:calcium_and_behavior}
    \end{subfigure}

    \caption{\textbf{Use of NLP4Neuro to predict behavioral output from simultaneous neural activity in larval zebrafish} (a) Volumetric two-photon calcium imaging of head-fixed, tail-freed larval zebrafish during visual stimulation while recording tail movements at 200 Hz. Pt: pretectum; nMLF: nucleus of the medial longitudinal fasciculus; Hb: hindbrain. (b) The NLP4Neuro pipeline selects an LLM to learn a mapping of GCaMP6s fluorescence images, recorded for $s$ (sequence length) frames, onto points along the fish tail, i.e. $\theta_i$ for $i \in [0,5]$. Each $\Delta \theta_i$ is re-set to a baseline of 0 (radians).}
    \label{fig:data_acqu}
    \end{figure}

\else
    \begin{figure}[!tbp]
    \centering
    \begin{subfigure}[b]{0.275\linewidth}
        \centering
        \raisebox{1.25mm}[0pt][0pt]{%
        \includegraphics[width=1\linewidth]{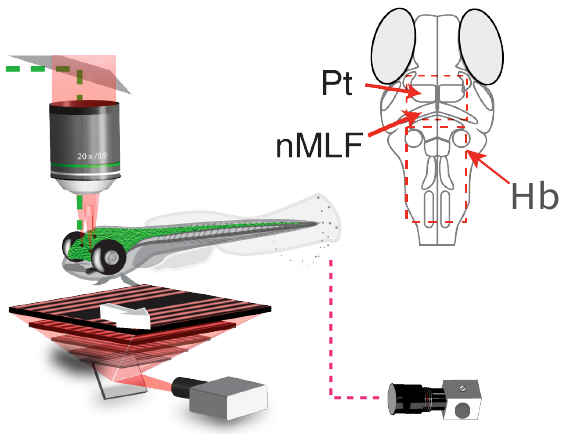}
        }
        \caption{}
        \label{fig:expsetup}
    \end{subfigure}
    \hspace{1cm}
    \begin{subfigure}[b]{0.6\linewidth}
        \centering
        \includegraphics[width=1\linewidth]{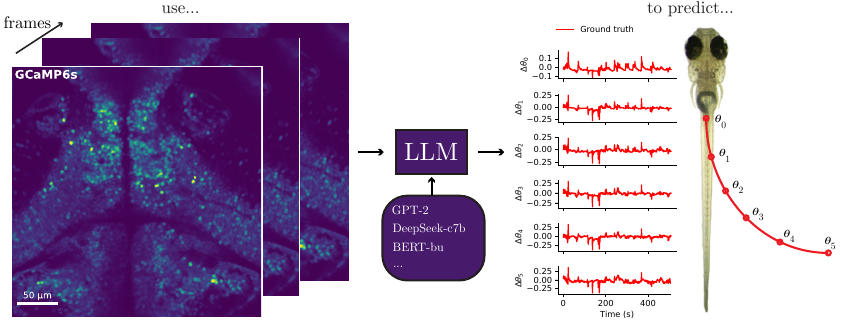}
        \caption{}
        \label{fig:calcium_and_behavior}
    \end{subfigure}

    \caption{\textbf{Use of NLP4Neuro to predict behavioral output from simultaneous neural activity in larval zebrafish} (a) Volumetric two-photon calcium imaging of head-fixed, tail-freed larval zebrafish during visual stimulation while recording tail movements at 200 Hz. Pt: pretectum; nMLF: nucleus of the medial longitudinal fasciculus; Hb: hindbrain. (b) The NLP4Neuro pipeline selects an LLM to learn a mapping of GCaMP6s fluorescence images, recorded for $s$ (sequence length) frames, onto points along the fish tail, i.e. $\theta_i$ for $i \in [0,5]$. Each $\Delta \theta_i$ is re-set to a baseline of 0 (radians).}
    \label{fig:data_acqu}
    \end{figure}
\fi
% ------------------------------

Tail movements were recorded at 200 Hz, wherein we tracked 6 tail angle segments \cite{Stih2019-uv}. Following two-photon image acquisition, uncompressed image stacks were corrected for motion artifacts using \textit{CaImAn} \cite{Giovannucci2019-ix}, an open-source calcium imaging processing library, and its implementation of the NoRMCorre algorithm, which calculates and aligns motion vectors with subpixel resolution. Using \textit{Suite2P} \cite{Pachitariu2016-vp} source identification and signal extraction, all further analyses utilized estimated fluorescence traces of signal sources presumed to represent the neural activity of single neurons. For simplicity, we refer to these extracted sources as neurons. 

\subsection{Models}
\textbf{Recurrent neural networks}\hspace{.5cm}
We consider two model classes of recurrent neural networks (RNNs). The discrete-time reservoir computer (RC) is a non-linear RNN with simple linear training. It relies on a sufficient richness of its reservoir layer to find coefficients for reservoir neurons which reconstruct a ground truth output time series. The Long Short-Term Memory network (LSTM) is a contemporary RNN which has been used in neural decoding, e.g. to delineate behavior \cite{Huang2020-tt} and for interacting with computers via brain machine interfaces \cite{Shevchenko2024-ta}.

\textbf{Transformer neural networks}\hspace{.5cm} We broadly sample from transformer model classes, including an encoder-based transformer, Bidirectional Encoder Representations from Transformers (BERT); a decoder-based transformer, Generative Pre-trained Transformer 2 (GPT-2); and a Mixture-of-Experts (MoE) model, DeepSeek-c7b. Unless otherwise specified, we use the pre-trained versions of each model, BERT-base-uncased (BERT-bu), GPT-2, and DeepSeek-c7b. Untrained versions copy only the architecture of these models.

\textbf{Mixture of Experts}\hspace{0.5cm} Mixture of Experts (MoE) models modify the decoder-based transformer through the application of a routing network, $(W_r,b_r)$, which computes scores or gates, given a hidden state $h(t)$. For DeepSeek-c7b $h(t) \in \mathbb{R}^{4096}$ is a context-enhanced token representation, similar to $z(t)$ in Eq. \ref{eq:attnfinal}. In DeepSeek-c7b's MoE architecture, as described in \cite{Guo2024-dk}, the input token $u(t)$ is linearly mapped into the 4096-dimensional space, then pushed through a transformer block, consisting of Eqs. \ref{eq:qkv}-\ref{eq:layernorm}. Then, a softmax layer is applied to a learnable routing matrix $W_r$, given bias $b_r$. The model is called a \textit{mixture of experts}, because it then recruits the top-$k$ experts ($e_i$) with indices $i$ (see Eqs. \ref{eq:routing}- \ref{eq:moe}). Finally, this output is projected back to either the token dimensionality or that required for a subsequent layer, using $W_{out}z(t)$. This comprises the updated token. 

\subsection{Input embeddings}\label{ssec:embeddings}
We consider five input embedding strategies, to delineate whether neural activity vectors as tokens can be transformed in a way that improves the performance of the model, in this case to predict tail component positions across frames for a sample of sequences. The strategies employed include: no strategy, or a linear mapping of inputs to the first transformer block; positional encoding of each token in relation to each other token \cite{Vaswani2017-nt}; relative positional encoding, which extends positional encoding to consider pairwise positions \cite{Shaw2018-jo}; and dimensionality reduction techniques, including Laplacian eigenmaps and sparse autoencoders \cite{Belkin2003-ry, Cunningham2023-ap, Ng2011Sparse}. Discussions of these methods are provided in the appendix.

\subsection{Evaluation}
We collected neural data from recordings of $N$ neurons in the pretectum, nucleus of the medial fasciculus, and hindbrain (see Fig. \ref{fig:data_acqu}) for $t$ discrete time steps. To align tail kinematics data that were simultaneously collected at a much higher sampling frequency (200 Hz), compared to the calcium imaging frequency of 1.1 Hz, we compressed this sample set from $\sim$500,000 steps to $\sim$4,000 steps (i.e. with small differences across fish) -- to match the neural data for sequence-to-sequence learning -- by simply averaging across all output time steps in a given frame. Tail data are arranged into a 6-panel time series, where each panel denotes a tail component, $\theta_i$, $i \in [0,5]$. 

We fine-tuned pre-trained models using a small number of training epochs, between 5 and 10 (10 epochs for Figs. \ref{fig:compare1} and \ref{fig:embeddings_all}-\ref{fig:saliency_overlays}, 5 for Fig. \ref{fig:pretrained_vs_untrained} and Table \ref{tab:all_results}). We used a batch size of 16 for Experiment 1 and 4 and 32 for Experiments 2-3, a learning rate of $1E-06$ for Experiment 1 and 4 and $1E-04$ for Experiments 2-3, and a train-validate-test split of 70:15:15 for Experiment 1 and 4, and 70:10:20 for Experiments 2-3. For all DeepSeek-c7b models, we used 8 experts, and a top $k$ of 2.

All models were built in PyTorch \cite{Paszke2019-pq}. The reservoir computer model was built manually in PyTorch using Eqs. \ref{eqn: rc1}-\ref{eqn: rc2}. We used built-in architectures and model weights for other models. For the LSTM, we inherited PyTorch's implementation with a single layer. For all large language models, we used weights and architecture from the transformers library (\href{https://pypi.org/project/transformers/}{https://pypi.org/project/transformers/}). All RNNs use the Adam optimizer \cite{Kingma2014-zp}; all transformers use the AdamW optimizer \cite{Loshchilov2017-hk}.

Each sequence is a sample in our training, validation, and test sets. To evaluate the performance of each model on its corresponding held-out test set, we computed the root-mean squared error (RMSE), as shown in Eq. \ref{eqn:rmse} between ground truth tail signal and predicted tail signal, across all sequences of sequence length $s$. The sum of RMSE values across all sequences is the metric reported for a given model type and sequence length. In addition to RMSE scores, to get an intuition for the fit of the sequence window-averaged test set performance, where for overlapping window time steps we averaged the predictions, we also looked at Pearson's product-moment correlation coefficient, $r$, between model tail sum predictions $\theta_{\text{sum},p}$ (see appendix) and ground truth tail sum data $\theta_{\text{sum},p}$. All reported error bars, unless otherwise stated, are the standard error of the mean.

\subsection{Salience mapping}\label{ssec:saliency}

Using our best-performing LLM, given a sequence length $s$, we computed the salience for a neuron $i$, at frame $t$, in sliding window $m$ -- where $m$ is the current sequence-to-sequence index -- of length $s$, corresponding to tail component predictions $\hat{\theta}_k$, $k \in \{0,\dots,5\}$, as an input-gradient score \cite{Simonyan2013-lu, Hou2023-ls}: 

\begin{equation}\label{eqn:saliency3}
  s^{(m)}_{i,t} \;=\;
  \left|\,x^{(m)}_{i,t}\,
       \frac{\partial f^{(m)}}{\partial x^{(m)}_{i,t}}
  \right|,
  \qquad
  f^{(m)} \;=\;
  \sum_{\tau,\,k=0}^{5}\hat\theta^{(m)}_{\tau,k},   
\end{equation}

where $f^{(m)}$ is a scalar value, and $\tau$ is a local frame counter within the sliding window $m$. We take the absolute value as we are only concerned with the magnitude of a neuron's impact on output frames.

\textit{For moment-to-moment salience}, $S_{i,t}$, to denote the contribution of a neuron to a particular frame of predicted output, we averaged this score for all overlapping windows containing the frame. 

For \textit{global salience}, $S_i$, we computed a single salience score for each neuron by averaging moment-to-moment salience scores over the full recording of $T$ frames. See the appendix and Eq. \ref{eqn:saliency}.

\section{Results}
\subsection{Experiment 1: Sequence-to-sequence learning of tail behavior from neural activity}

From Fig. \ref{fig:compare1}, we observe that the reservoir computer is ill-suited for sequence-to-sequence learning of neural data compared to other models. It performs the worst, and yields the highest RMSE score compared to all other models. Significance for this experiment is computed using a Mann-Whitney U test. The LSTM, conversely, performs well on a range of sequence lengths, whilst using only a small percentage of the computational footprint of the transformer models considered. 

\iflatexml
    \begin{figure}[!tbp]
      \centering
        \begin{subfigure}[t]{\textwidth}
          \centering
          \includegraphics[width=1\textwidth]{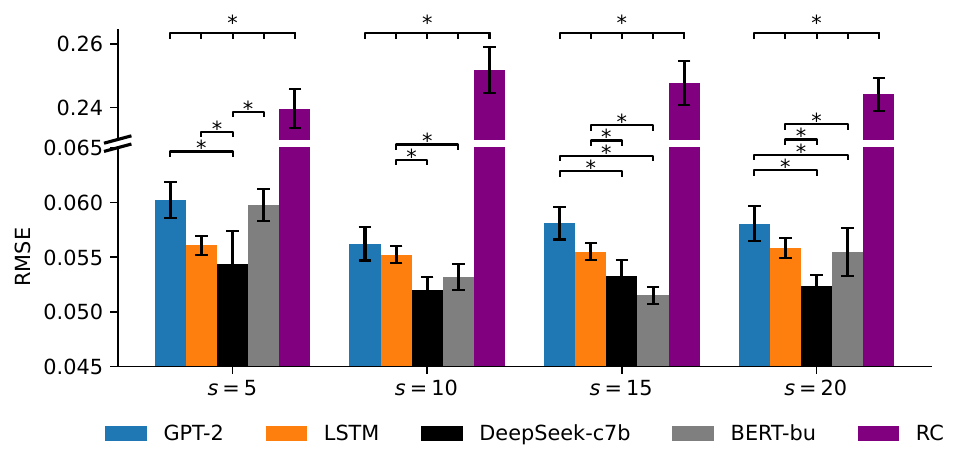}
          \caption{}
          \label{fig:compare1}
        \end{subfigure}\\[1ex]
        \begin{subfigure}[t]{1\textwidth}
          \centering
          \includegraphics[width=\textwidth]{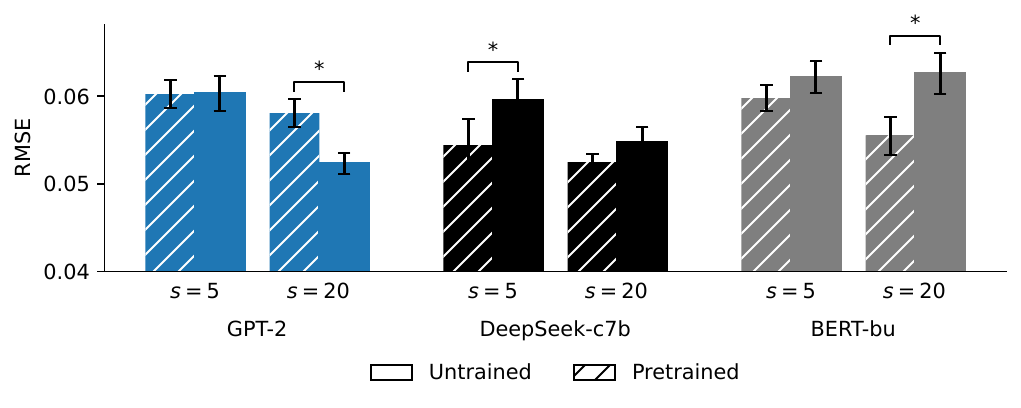}
          \caption{}
          \label{fig:pretrained_vs_untrained}
        \end{subfigure}\\[1ex]
        \begin{subfigure}[b]{\textwidth}
          \centering
          \includegraphics[width=\textwidth]{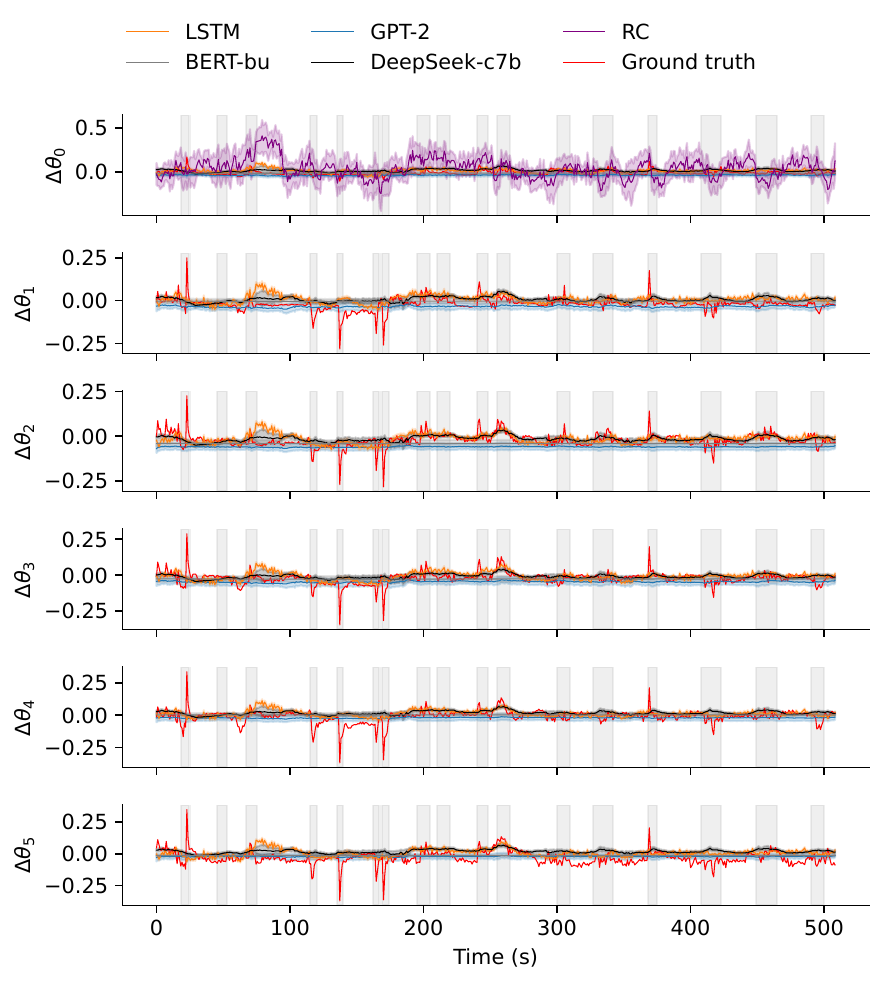}
          \caption{}
          \label{fig:allmodelpreds}
        \end{subfigure}
      \caption{\textbf{Pre-trained DeepSeek-c7b captures tail kinematic events, simple RNNs do not.}\\ (a) Grouped model test set performance across fish, with increasing sequence lengths, $s$. (b) Pre-trained versus un-trained performance for transformer models. For (a-b), we use data from $N=5$ fish. A * denotes significant differences ($p<0.05$) using the Mann-Whitney U test. (c) Sampled behavioral output and test set predictions with 95\% confidence intervals, averaged across overlapping prediction window frames, for one fish, with $s=10$. The RC is shown only in the panel for $\Delta\theta_0$.}
      \label{fig:resultsexp1_2}
    \end{figure}
\else
    \begin{figure}[!tbp]
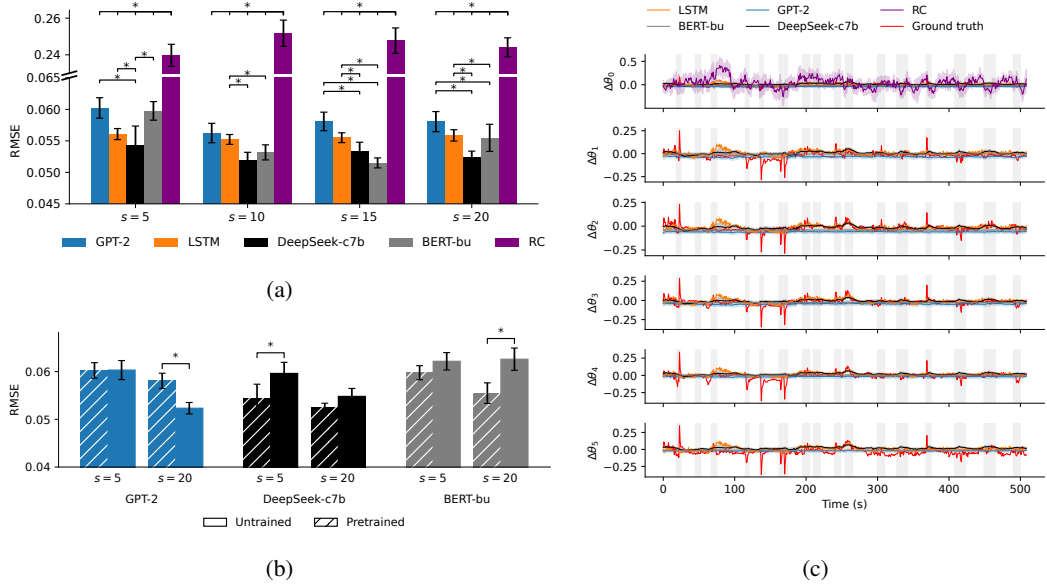

      \centering
      % Left column: (a) above (c)
      \begin{minipage}[b]{0.53\textwidth}
        \centering
        % Subfigure (a)
        \begin{subfigure}[b]{\textwidth}
          \centering
          \includegraphics[width=1\textwidth]{figures/fig2a.pdf}
          \caption{}
          \label{fig:compare1}
        \end{subfigure}
        \\[1ex]
        % Subfigure (c)
        \begin{subfigure}[b]{1\textwidth}
          \centering
          \includegraphics[width=\textwidth]{figures/fig2b.pdf}
          \caption{}
          \label{fig:pretrained_vs_untrained}
        \end{subfigure}
      \end{minipage}\hfill
      % Right column: (b)
      \begin{minipage}[b]{0.45\textwidth}
        \centering
        \begin{subfigure}[b]{\textwidth}
          \centering
          \raisebox{3mm}[0pt][0pt]{%
          \includegraphics[width=\textwidth]{figures/fig2c.pdf}
          }
          \caption{}
          \label{fig:allmodelpreds}
        \end{subfigure}
      \end{minipage}
      \caption{\textbf{Pre-trained DeepSeek-c7b captures tail kinematic events, simple RNNs do not.}\\ (a) Grouped model test set performance across fish, with increasing sequence lengths, $s$. (b) Pre-trained versus un-trained performance for transformer models. For (a-b), we use data from $N=5$ fish. A * denotes significant differences ($p<0.05$) using the Mann-Whitney U test. (c) Sampled behavioral output and test set predictions with 95\% confidence intervals, averaged across overlapping prediction window frames, for one fish, with $s=10$. The RC is shown only in the panel for $\Delta\theta_0$.}
      \label{fig:resultsexp1_2}
    \end{figure}
\fi
    
Among transformer neural networks, GPT-2 produces the highest (poorest) RMSE scores across fish and train-validate trials, but still fares better than the reservoir computer. BERT-bu achieves better (lower) RMSE scores than GPT-2, but does not appear to track well with the data (Fig. \ref{fig:allmodelpreds}, Table \ref{tab:all_results}). DeepSeek-c7b best predicts tail component behavior from neural data, achieving the lowest RMSE scores for all sequence lengths except $s=15$, and having the highest Pearson's $r$-value (Table \ref{tab:all_results}). Looking at DeepSeek-c7b's performance across sequence lengths, compared to the best RNN (the LSTM), it is apparent that the margin of improvement for the MoE model is greater at longer sequence lengths, for example comparing $s=5$ to $s=20$. This is not the case for the LSTM, which actually performs best compared to all other models at $s=5$.

We observe in Fig. \ref{fig:allmodelpreds} that DeepSeek-c7b captures tail events better than it does tail amplitudes. When grouping events into bursts of tail activity, it, for example, captures the timing of such events, but not always their valence or value. We suspect that this is a result of short, high-amplitude bursts not being sufficiently penalized by the RMSE loss function. The LSTM does not appear to capture these tail events as well as DeepSeek-c7b, but does better following the amplitude and timing of events compared to other transformers. We see also that other transformers, GPT-2 and BERT-bu, appear to undershoot the amplitude and frequency of the time-varying tail signal, although their predictions are centered at roughly the ground truth baseline. The RNNs considered, including both the reservoir computer and LSTM, appear to overshoot the frequency, where the RC most harshly exaggerates the predicted amplitude. 

In Table \ref{tab:all_results}, we see that Pearson's product-moment correlation coefficient, $r$, captures a similar trend compared to the RMSE results. Namely, that DeepSeek-c7b most closely matches ground truth tail kinematics test set data ($r=0.08$ to $0.1$) for all $s$, followed by the LSTM, with a weak positive correlation coefficient of $r=0.07$. The reservoir computer is, similarly, least correlated to the ground truth in comparison to the other models.

\subsection{Experiment 2: Comparison of pre-trained language models to un-trained models}

Considering LLMs only, we ask whether using the pre-trained weights from these off-the-shelf models informs improved predictions after fine-tuning for a small number of epochs on neural activity data for predicting tail behavior. As shown in Fig. \ref{fig:pretrained_vs_untrained}, using sequence lengths $s=5$ and $s=20$ frames, we train, validate, and test the pre-trained and un-trained variants of the model. Each pre-trained model loads its weights using the AutoModel transformers package, \href{https://pypi.org/project/transformers/}{https://pypi.org/project/transformers/}. Each un-trained model loads only the architecture, using the AutoConfig transformers package. 

For GPT-2, on smaller sequence lengths, pre-trained versus untrained variants are not significantly different. On the test set for $s=20$, the un-trained GPT-2 model actually outperforms its pre-trained variant. Using DeepSeek-c7b, we find that for both sequence lengths, the pre-trained model weights inform improvements in performance. This trend is also observed with BERT-bu, particularly for the longer sequence length $s=20$.

\begin{table*}[!htbp]
\centering
\caption{Results Summary for Experiment 1: Performance vs.\ Sequence Length. Each model lists the RMSE $\pm$~standard error and Pearson’s $r \pm$~standard error. \textbf{Bold} marks the lowest RMSE in each column. A $^{*}$ indicates $p<0.05$ (Mann-Whitney U test) for the LSTM vs. DeepSeek-c7b.}
\label{tab:all_results}
\begin{tabular}{llcccc}
\toprule
\textbf{Model} &  & $s=5$ & $s=10$ & $s=15$ & $s=20$ \\
\midrule
\multirow{2}{*}{BERT-bu} 
    & RMSE & 0.060 $\pm$ 0.001 & 0.053 $\pm$ 0.001 & \textbf{0.052 $\pm$ 0.001} & 0.055 $\pm$ 0.002 \\
    & $r$  & 0.05  $\pm$ 0.01  & 0.06  $\pm$ 0.01  & 0.06  $\pm$ 0.01           & 0.06 $\pm$ 0.01 \\
\addlinespace[0.3ex]
\multirow{2}{*}{DeepSeek-c7b} 
    & RMSE & \textbf{0.054 $\pm$ 0.003}$^{*}$ & \textbf{0.052 $\pm$ 0.001}$^{*}$ & 0.053 $\pm$ 0.001$^{*}$ & \textbf{0.052 $\pm$ 0.001}$^{*}$ \\
    & $r$  & 0.10 $\pm$ 0.01 & 0.08 $\pm$ 0.01 & 0.10 $\pm$ 0.01 & 0.09 $\pm$ 0.01 \\
\addlinespace[0.3ex]
\multirow{2}{*}{GPT-2} 
    & RMSE & 0.060 $\pm$ 0.002 & 0.056 $\pm$ 0.002 & 0.058 $\pm$ 0.001 & 0.058 $\pm$ 0.002 \\
    & $r$  & 0.07 $\pm$ 0.01 & 0.06 $\pm$ 0.01 & 0.05 $\pm$ 0.01 & 0.04 $\pm$ 0.01 \\
\addlinespace[0.3ex]
\multirow{2}{*}{LSTM} 
    & RMSE & 0.056 $\pm$ 0.001 & 0.055 $\pm$ 0.001 & 0.056 $\pm$ 0.001 & 0.056 $\pm$ 0.001 \\
    & $r$  & 0.07 $\pm$ 0.02 & 0.07 $\pm$ 0.02 & 0.06 $\pm$ 0.02 & 0.07 $\pm$ 0.02 \\
\addlinespace[0.3ex]
\multirow{2}{*}{RC} 
    & RMSE & 0.240 $\pm$ 0.006 & 0.252 $\pm$ 0.007 & 0.248 $\pm$ 0.007 & 0.244 $\pm$ 0.005 \\
    & $r$  & 0.01 $\pm$ 0.01 & 0.02 $\pm$ 0.01 & 0.00 $\pm$ 0.01 & 0.01 $\pm$ 0.01 \\
\bottomrule
\end{tabular}
\end{table*}

\subsection{Experiment 3: Compare encodings of neural activity data}

We consider five input encoding strategies, to explore whether embedding input tokens with an encoding scheme helps to transfer the task of reading temporal dependencies between neural data into a representation which is amenable to pre-trained language models. 

\begin{figure}[!tbp]
    \hspace{-.7cm}
    \centering
    \includegraphics[width=.9\textwidth]{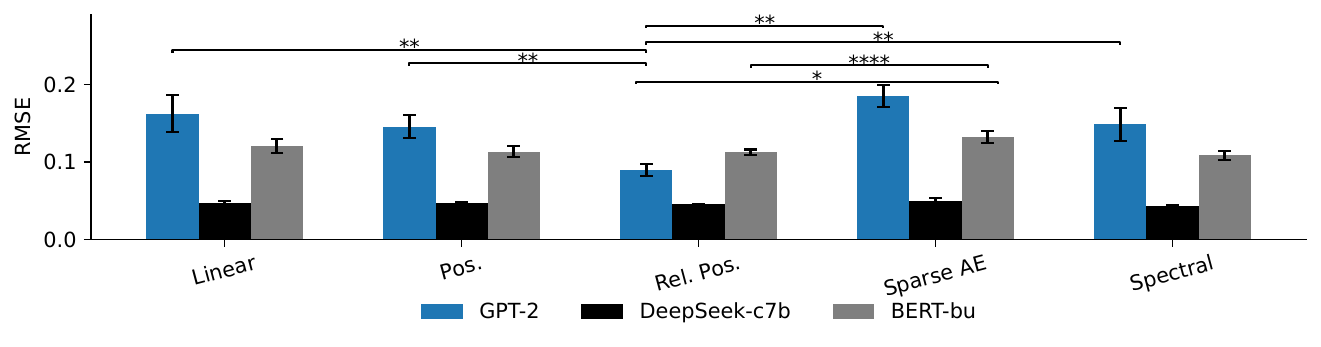}
    \caption{\textbf{Relative position improves performance across transformers. DeepSeek-c7b is not sensitive to embedding strategies.} Performance is shown for a representative fish, with $s=20$. Differences (*) across model groups, and within groups (**-**** for GPT-2, DeepSeek, and BERT-bu respectively) are shown for each embedding strategy, using a Mann-Whitney U test ($p < 0.05$).}
    \label{fig:embeddings_all}
\end{figure}

See Fig. \ref{fig:embeddings_all}. Since neural data tokens can be simply mapped onto the first transformer block with a linear mapping, we consider this as a null encoding (Linear), and add reasonable extensions to this, such as the encoding of the position of each token (Pos.), each token's position in relation to others (Rel. Pos.), and dimensionally-reduced neural token representations via network-based and neural network-based embedding techniques: Laplacian eigenmaps (i.e. Spectral embedding, or Spectral) and sparse autoencoders (Sparse AE). 

For a sequence length $s=20$ and representative fish, we compare the embedding strategies across models, and also within model classes. In Fig. \ref{fig:embeddings_all}, we observe that relative position encoding (Rel. Pos.) improves the performance across all transformer models broadly, in particular because it significantly improves the performance of GPT-2 on the test data. We find significant differences in performance between relative positioning and all other strategies for GPT-2, by the Mann-Whitney U test. Comparing DeepSeek-c7b on all embedding strategies, we find no significant difference in performance, although we do see a strong improvement in performance compared to other models. One possible explanation here could be that, for DeepSeek-c7b only, the experts in each MoE layer contribute disproportionately more to model predictions than earlier transformer blocks.

\subsection{Experiment 4: Use the best model to delineate neural circuitry}

Considering DeepSeek-c7b, which has achieved the best performance of the models considered, in terms of low RMSE and high correlation to the test set data, we use both global and moment-to-moment salience mapping to decode the relative contribution of neurons across neural data tokens for predicting motor outputs.

To test DeepSeek-c7b's capacity to retrieve task critical neurons, we extracted the top $k$ most salient input features $S_i$ from individual fish, and marked their spatial location and importance (as the dot size). In Fig. \ref{fig:fish11_overlay} ($k=20$), we see that the spatial locations of the most salient neurons appears to be mostly focused in the Pt, which encodes visual motion and is implicated in tail behavior \cite{Antinucci2019-wy}; this observation is limited, however, in that all neural data is visually evoked, not spontaneous. All cells are clustered into their respective region in Fig. \ref{fig:fish11_clusters}. We validate our intuition of the observed data in Fig. \ref{fig:fish11_importance}, i.e. the mean salience of top neurons ($k=10$) in the Pt is greater than in the Hb across fish, using data from three fish. Here we measure significant differences with a Student's t-test ($p < 0.05$). 

To demonstrate DeepSeek-c7b's abilities as a tool for uncovering neural circuitry, we first evaluated the model's capacity to capture left and right tail movements, referring to well-known visuomotor regions, as documented in \cite{Naumann2016-mx}. Using the tail sum, we plot examples of successful predictions for a representative zebrafish in Fig. \ref{fig:fish12_concat}. Our results show that DeepSeek-c7b's predictions for this fish are biased towards and better at modeling right tail movements. Using global salience, we see in Fig. \ref{fig:fish12_saliency} that there is asymmetry between right tail motion-predicting neurons and left-tail motion predicting neurons. However, we also note here that neurons which are predictive of such tail motions may contribute, for example, with lack of activity rather than high activity, which we are not testing for.

In Figs. \ref{fig:seq75}-\ref{fig:seq75bars}, we highlight the power of using moment-to-moment salience, $S_{i,t}$ (See Ssec. \ref{ssec:saliency}), to capture neural contributions to behavior. A positive exemplar sequence is shown in Fig. \ref{fig:seq75}, where DeepSeek-c7b generally anticipates tail behavior from neural activity data. Corresponding to frame 77 in this sequence, we are able to see in Fig. \ref{fig:seq75rois}-\ref{fig:seq75bars} that the right side of the midbrain contributes most to right tail predictions, and that most contributing neurons are in the Pt. Both observations appear to validate that the right pretectum is involved in controlling rightward tail movements \cite{Naumann2016-mx}. 

\iflatexml
    \begin{figure}[!h] 
        \centering
        \begin{subfigure}[t]{0.275\textwidth}
            \centering
            \includegraphics[width=.65\textwidth]{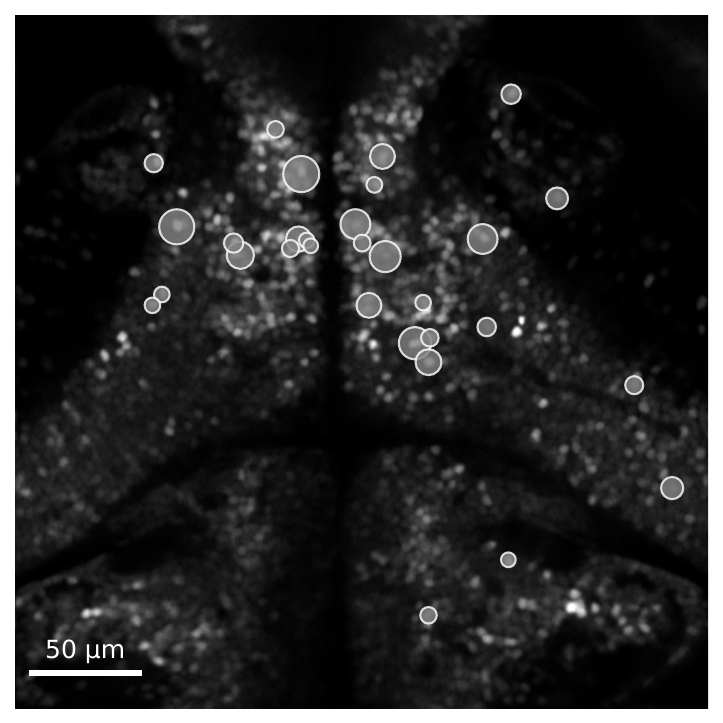}
            \caption{}
            \label{fig:fish11_overlay}
        \end{subfigure}
        % \hspace*{3mm}
        \begin{subfigure}[t]{0.275\textwidth}
            \centering
            \includegraphics[width=.65\textwidth]{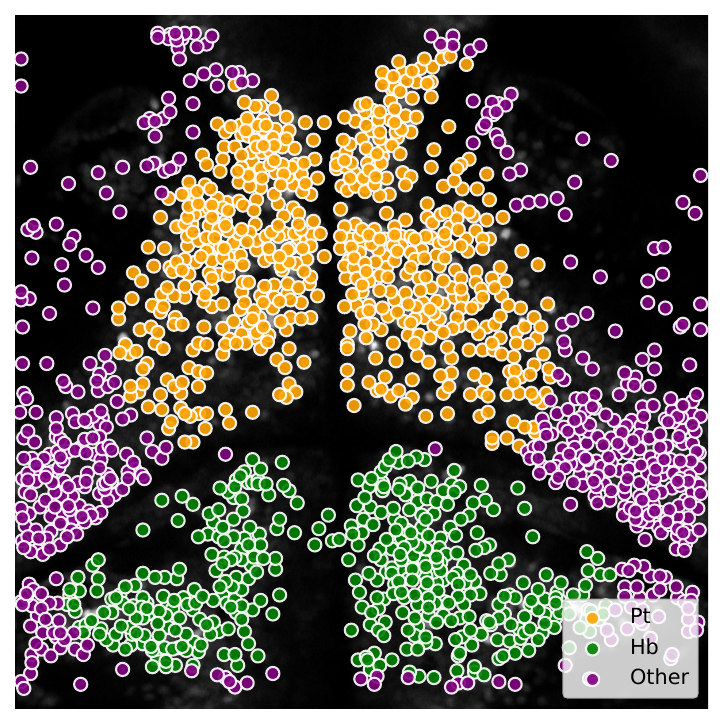}
            \caption{}
            \label{fig:fish11_clusters}
        \end{subfigure}
        % \hfill
        \begin{subfigure}[t]{0.3\textwidth}
            \centering
            \includegraphics[width=.73\textwidth]{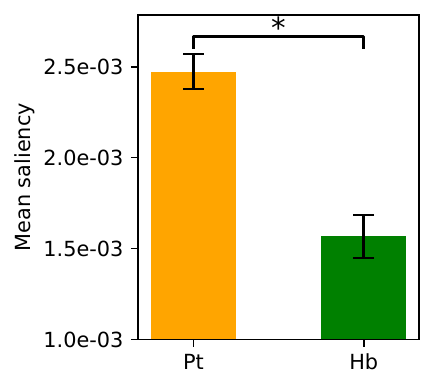}
            \caption{}
            \label{fig:fish11_importance}
        \end{subfigure}
    
        \caption{\textbf{Globally salient neurons are found in the pretectum (Pt)}. (a) Neuron salience for DeepSeek-c7b, projected onto the GCaMP6s image for a representative fish. Dot size reflects salience score for predicting tail output. (b) Regional clustering of all neurons. (c) Mean scores from $k=10$ most salient neurons per region across 3 fish. Significance is measured using a Student's t-test.}
        \label{fig:saliency_overlays}
    \end{figure}
\else
    \begin{figure}[!h] 
        \centering
        \begin{subfigure}[b]{0.275\textwidth}
            \centering
            \includegraphics[width=.65\textwidth]{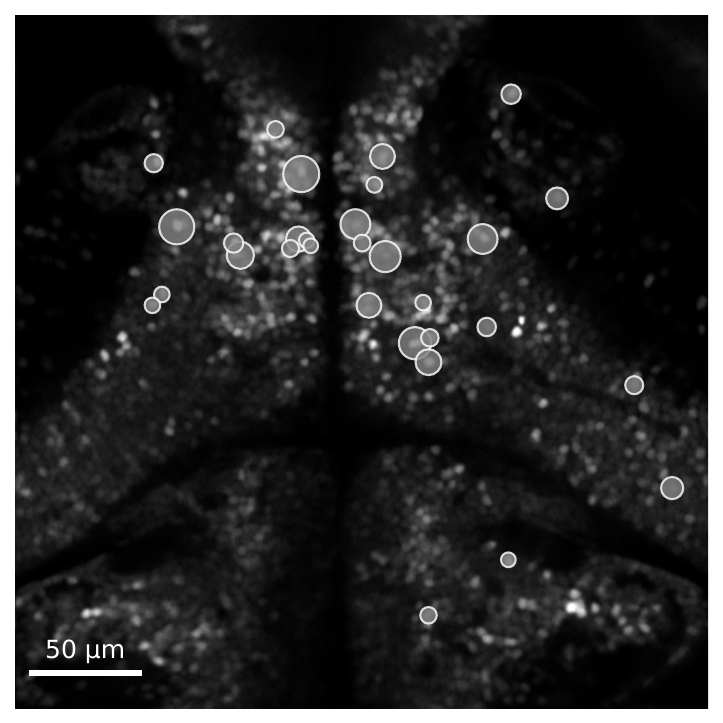}
            \caption{}
            \label{fig:fish11_overlay}
        \end{subfigure}\hspace{.25cm}
        % \hspace*{3mm}
        \begin{subfigure}[b]{0.275\textwidth}
            \centering
            \includegraphics[width=.65\textwidth]{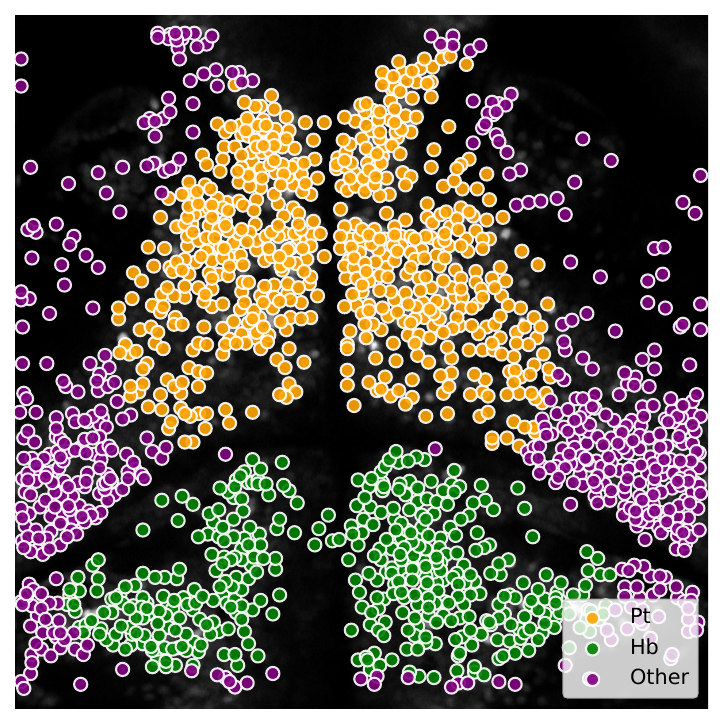}
            \caption{}
            \label{fig:fish11_clusters}
        \end{subfigure}
        % \hfill
        \begin{subfigure}[b]{0.3\textwidth}
            \centering
            \hspace*{-8mm}
            \raisebox{-2.3mm}[0pt][0pt]{%
            \includegraphics[width=.73\textwidth]{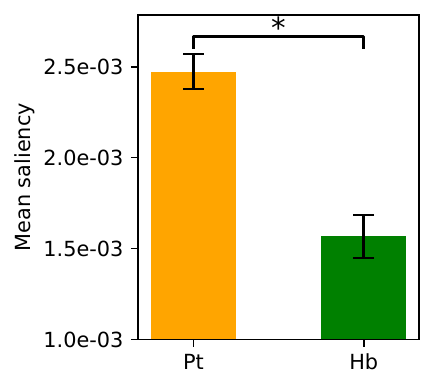}
            }
            \caption{}
            \label{fig:fish11_importance}
        \end{subfigure}
    
        \caption{\textbf{Globally salient neurons are found in the pretectum (Pt)}. (a) Neuron salience for DeepSeek-c7b, projected onto the GCaMP6s image for a representative fish. Dot size reflects salience score for predicting tail output. (b) Regional clustering of all neurons. (c) Mean scores from $k=10$ most salient neurons per region across 3 fish. Significance is measured using a Student's t-test.}
        \label{fig:saliency_overlays}
    \end{figure}
\fi

% ISSUE -> b is massive...a is tiny
\iflatexml
    \begin{figure}[!h]
        \centering
        \begin{subfigure}[t]{0.75\textwidth}
            \centering
            \includegraphics[width=\textwidth]{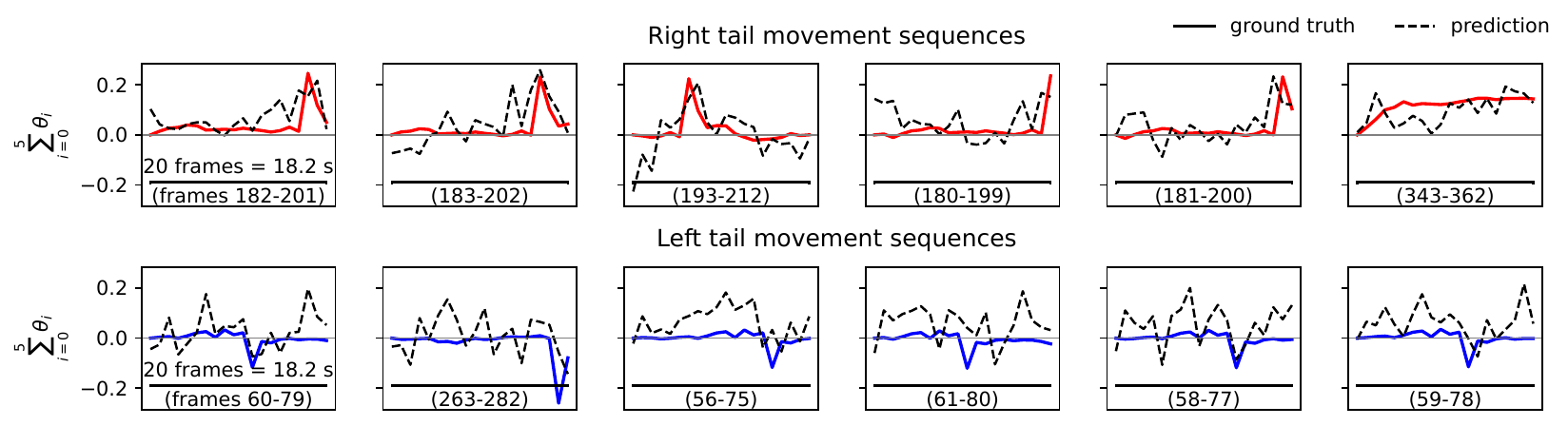}
            \caption{}
            \label{fig:fish12_concat}
        \end{subfigure}\\[1ex]
    \begin{subfigure}[t]{0.275\textwidth}
        \centering
    \end{subfigure}
    \begin{subfigure}[t]{0.25\textwidth}
        \centering
        \includegraphics[width=.75\textwidth]{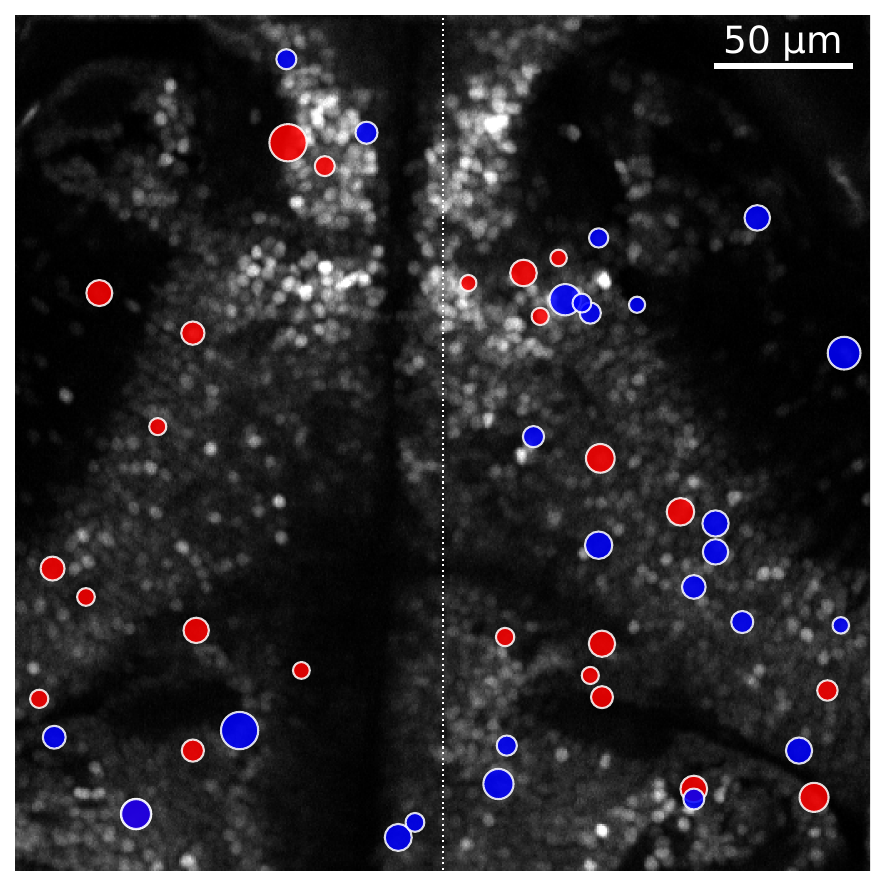}  
        \caption{}  % this is (b)
        \label{fig:fish12_saliency}
    \end{subfigure}
    \begin{subfigure}[t]{0.275\textwidth}
        \centering
    \end{subfigure}    
        \caption{\textbf{DeepSeek predicts rightward and leftward tail movements, and is interpretable}. Salient neurons show bilateral symmetry. (a) Predicted and ground truth sequences ($s=20$) for right (red) and left (blue) tail movements. (b) Left (blue) and right (red) motion salience overlay for a representative zebrafish. Larger dot size denotes a higher relative salience score.}
        \label{fig:saliency_overlays}
    \end{figure}
\else
    \begin{figure}[!h]
    \hspace{-.5cm}
        \centering
        % (a) takes 60% of line width
        \begin{subfigure}[b]{0.75\textwidth}
            \centering
            \includegraphics[width=\textwidth]{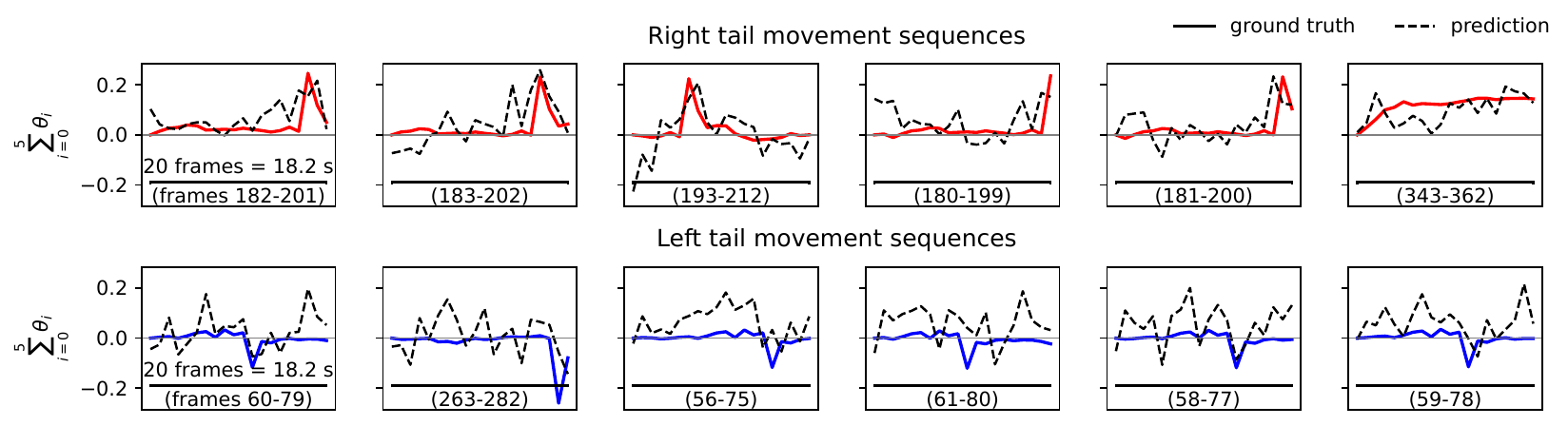}
            \caption{}
            \label{fig:fish12_concat}
        \end{subfigure}\hfill
    \begin{subfigure}[b]{0.25\textwidth}
        \centering
        \includegraphics[width=.75\textwidth]{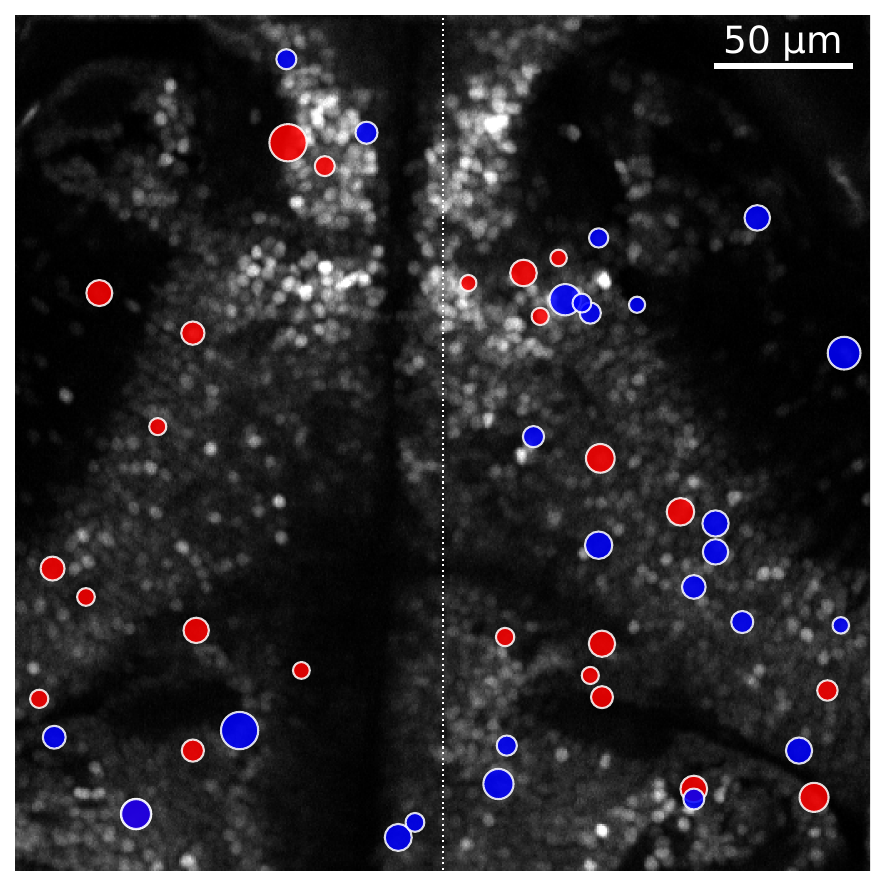}
        \caption{}  % this is (b)
        \label{fig:fish12_saliency}
    \end{subfigure}
    
        \caption{\textbf{DeepSeek predicts rightward and leftward tail movements, and is interpretable}. Salient neurons show bilateral symmetry. (a) Predicted and ground truth sequences ($s=20$) for right (red) and left (blue) tail movements. (b) Left (blue) and right (red) motion salience overlay for a representative zebrafish. Larger dot size denotes a higher relative salience score.}
        \label{fig:saliency_overlays}
    \end{figure}
\fi

% Issue.. (a) should be lower...
\iflatexml
    \begin{figure}[!h]
    \centering
    \begin{subfigure}[t]{0.32\textwidth}
        \centering
        \includegraphics[width=.85\textwidth]{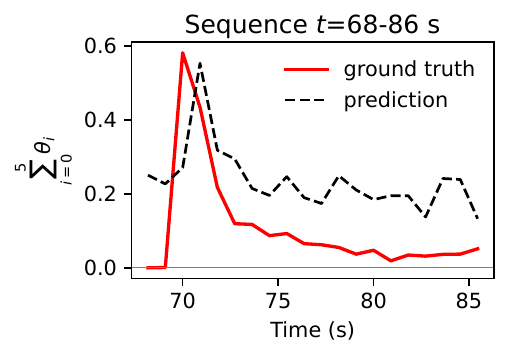}%
        \caption{}
        \label{fig:seq75}
    \end{subfigure}
    \begin{subfigure}[t]{0.3\textwidth}
        \centering
        \includegraphics[width=2cm]{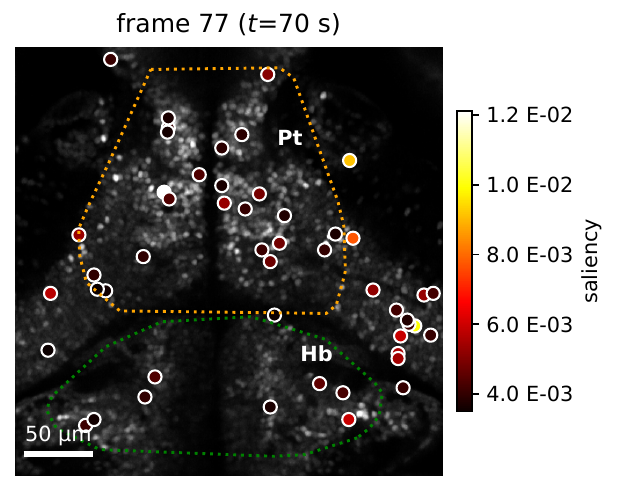}%
        \caption{}
        \label{fig:seq75rois}
    \end{subfigure}
    \begin{subfigure}[t]{0.33\textwidth}
        \centering
        \includegraphics[width=\textwidth]{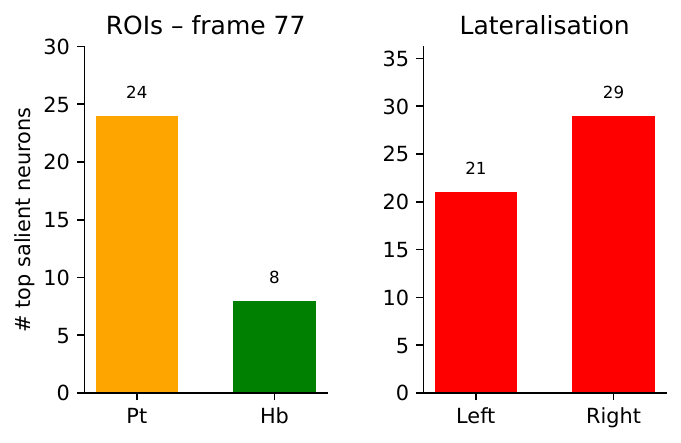}
        \caption{}
        \label{fig:seq75bars}
    \end{subfigure}    
    \caption{\textbf{Moment-to-moment salience mapping identifies known functional regions and  lateralisation}. (a) An exemplar sequence with predicted right tail movements from DeepSeek-c7b ($s=20$). (b) Spatial  distribution of the top 50 most salient pretectal (Pt), hindbrain (Hb), and other neurons for the target frame, corresponding to peak frame 77. (c) Deepseek-c7b correctly predicts highly salient neurons in the right hemisphere for a rightward directed tail movement for this frame.}
        \label{fig:saliency_overlays}
    \end{figure}
\else
    \begin{figure}[!h]
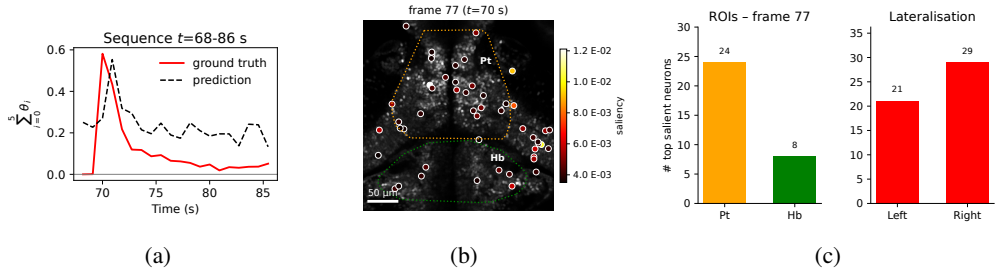

        \centering
          \begin{subfigure}[t]{0.32\textwidth}
            \centering
            \hspace*{-5mm}
            \raisebox{.6mm}[0pt][0pt]{%
            \includegraphics[width=.85\textwidth]{figures/fig6a.pdf}%
            }
            \caption{}
            \label{fig:seq75}
          \end{subfigure}
          % \hfill
          \begin{subfigure}[t]{0.3\textwidth}
            \centering
            \hspace*{0mm}
            \raisebox{1.5mm}[0pt][0pt]{%
            \includegraphics[width=.87\textwidth]{figures/fig6b.pdf}%
            }
            \captionsetup{margin={-7mm,0mm}}
            \caption{}
            \label{fig:seq75rois}
          \end{subfigure}
          % \hfill
          \begin{subfigure}[t]{0.33\textwidth}
            \centering
            \includegraphics[width=\textwidth]{figures/fig6c.pdf}
            \caption{}
            \label{fig:seq75bars}
          \end{subfigure}
        
            \caption{\textbf{Moment-to-moment salience mapping identifies known functional regions and  lateralisation}. (a) An exemplar sequence with predicted right tail movements from DeepSeek-c7b ($s=20$). (b) Spatial  distribution of the top 50 most salient pretectal (Pt), hindbrain (Hb), and other neurons for the target frame, corresponding to peak frame 77. (c) Deepseek-c7b correctly predicts highly salient neurons in the right hemisphere for a rightward directed tail movement for this frame.}
            \label{fig:saliency_overlays}
        \end{figure}
\fi

\newpage
\section{Conclusion}
We show that pre-trained, off-the-shelf large language models, and in particular, the Mixture of Experts-based large language model DeepSeek-c7b, are effective neural data learners, in particular when compared to RNNs. Surprisingly, we also find that the pre-trained weights for MoE models, which were learned over trillions of natural language tokens \cite{Guo2024-dk}, generalize well onto a neural activity decoding task. Additionally, we find that off-the-shelf token embedding strategies (for DeepSeek-c7b) are not sensitive to neural data tokens, suggesting that the transpose of each neural activity vector is likely as important as the time-varying nature of the population. Transforming this vector into a reduced feature space could simplify redundant neurons, but does not appear to improve the capacity of the model to delineate behavior. This result could also indicate that the experts featured in the MoE architecture are more prominently implicated in model predictions. 

We observe that both our global and moment-to-moment salience mapping approaches provide interpretable views of neurons which contribute to model tail predictions, and that these validate a known functional downstream pretectum-to-motor relationship \cite{Naumann2016-mx}. This is exciting for the reason that we are able to optogenetically photostimulate these neurons which most likley contribute to tail behavior, for the purpose of uncovering new functional connectivity pathways.

\section{Limitations}
We do not continue to fine tune models past 5 to 10 epochs of fine tuning, in order to: (1) retain the benefits we observe in the pre-trained weights; (2) reduce model training times, namely for DeepSeek-c7b; and (3) improve the accessibility of the findings. However, we also find that fine-tuning models past this number of epochs does not result in further improvements in performance. Expanding on (2), we use quantization (see appendix), which may disrupt LLM generalization \cite{Liu2023-yc}.

Embedding strategy results for DeepSeek-c7b suggest a dominance in the contribution of top-k Expert weights as opposed to weights in other transformer blocks. Follow-up work should evaluate the layer-wise distribution of weight changes after fine-tuning.

All observed neural responses are evoked, not spontaneous. Therefore, technically those neuron features which control model output behaviors can be only at-best recognized as routers for upstream OMR stimuli.  

Constructed salience maps show which neurons, as features, are controllers for predicted model activity, i.e. that which corresponds with animal behavior. However, this does not imply causality, and should be further validated with future photostimulation experiments.

\section{Broader impacts}
We present a surprising finding, that off-the-shelf MoE transformers, fine-tuned for a small number of epochs, are effective neural decoders. DeepSeek-c7b, as the best performing model, also has weights which are entirely accessible. Since, using NLP4Neuro, one can easily add state-of-the-art LLMs to our pipeline, together these provide an emerging opportunity for ML practitioners and Open science. 

Using our best model, we are able to determine the contribution of neurons (as model features) to zebrafish behavior. This tool will motivate future validation experiments with holographic photostimulation for mapping cause-effect functional connectivity, in service of improving brain-machine interfaces and modeling neurodegenerative disease.

\section{Code availability}
All code and data for this project are accessible at \href{https://github.com/Naumann-Lab/nlp4neuro.git}{https://github.com/Naumann-Lab/nlp4neuro.git}
\vspace{.5cm}

\newpage
\bibliography{neurips_2025}

\begin{thebibliography}{60}
\providecommand{\natexlab}[1]{#1}
\providecommand{\url}[1]{\texttt{#1}}
\expandafter\ifx\csname urlstyle\endcsname\relax
  \providecommand{\doi}[1]{doi: #1}\else
  \providecommand{\doi}{doi: \begingroup \urlstyle{rm}\Url}\fi

\bibitem[Unk()]{UnknownUnknown-yw}
Neural data transformer 3: A foundation model for motor cortical decoding.
\newblock \url{https://www.krellinst.org/csgf/conf/2024/abstracts/ye2022}.
\newblock Accessed: 2025-5-15.

\bibitem[Andalman et~al.(2019)Andalman, Burns, Lovett-Barron, Broxton, Poole, Yang, Grosenick, Lerner, Chen, Benster, Mourrain, Levoy, Rajan, and Deisseroth]{Andalman2019-dz}
Aaron~S Andalman, Vanessa~M Burns, Matthew Lovett-Barron, Michael Broxton, Ben Poole, Samuel~J Yang, Logan Grosenick, Talia~N Lerner, Ritchie Chen, Tyler Benster, Philippe Mourrain, Marc Levoy, Kanaka Rajan, and Karl Deisseroth.
\newblock Neuronal dynamics regulating brain and behavioral state transitions.
\newblock \emph{Cell}, 177\penalty0 (4):\penalty0 970--985.e20, May 2019.

\bibitem[Antinucci et~al.(2019)Antinucci, Folgueira, and Bianco]{Antinucci2019-wy}
Paride Antinucci, Mónica Folgueira, and Isaac~H Bianco.
\newblock Pretectal neurons control hunting behaviour.
\newblock \emph{Elife}, 8, October 2019.

\bibitem[Azabou et~al.(2024)Azabou, Pan, Arora, Knight, Dyer, and Richards]{Azabou2024-cu}
Mehdi Azabou, Krystal~Xuejing Pan, Vinam Arora, Ian~Jarratt Knight, Eva~L Dyer, and Blake~Aaron Richards.
\newblock Multi-session, multi-task neural decoding from distinct cell-types and brain regions.
\newblock In \emph{The Thirteenth International Conference on Learning Representations}, October 2024.

\bibitem[Barbu()]{BarbuUnknown-yd}
Andrei Barbu.
\newblock Population transformer.
\newblock \url{https://glchau.github.io/population-transformer/}.
\newblock Accessed: 2025-5-8.

\bibitem[Bardozzo et~al.(2024)Bardozzo, Terlizzi, Simoncini, Lió, and Tagliaferri]{Bardozzo2024-vm}
Francesco Bardozzo, Andrea Terlizzi, Claudio Simoncini, Pietro Lió, and Roberto Tagliaferri.
\newblock Elegans-{AI}: How the connectome of a living organism could model artificial neural networks.
\newblock \emph{Neurocomputing}, 584\penalty0 (127598):\penalty0 127598, June 2024.

\bibitem[Belkin and Niyogi(2003)]{Belkin2003-ry}
Mikhail Belkin and Partha Niyogi.
\newblock Laplacian eigenmaps for dimensionality reduction and data representation.
\newblock \emph{Neural Comput.}, 15\penalty0 (6):\penalty0 1373--1396, June 2003.

\bibitem[Buehler(2023)]{Buehler2023-wo}
Markus~J Buehler.
\newblock {MeLM}, a generative pretrained language modeling framework that solves forward and inverse mechanics problems.
\newblock \emph{J. Mech. Phys. Solids}, 181\penalty0 (105454):\penalty0 105454, December 2023.

\bibitem[Chau et~al.(2024)Chau, Wang, Talukder, Subramaniam, Soedarmadji, Yue, Katz, and Barbu]{Chau2024-uj}
Geeling Chau, Christopher Wang, Sabera Talukder, Vighnesh Subramaniam, Saraswati Soedarmadji, Yisong Yue, Boris Katz, and Andrei Barbu.
\newblock Population transformer: Learning population-level representations of neural activity.
\newblock \emph{arXiv [cs.LG]}, June 2024.

\bibitem[Chen et~al.(2024)Chen, Dumas, Watson, Vincoff, Peng, Zhao, Hong, Pertsemlidis, Shaepers-Cheu, Wang, Srijay, Monticello, Vure, Pulugurta, Kholina, Goel, DeLisa, Truant, Aguilar, and Chatterjee]{Chen2024-zf}
Tianlai Chen, Madeleine Dumas, Rio Watson, Sophia Vincoff, Christina Peng, Lin Zhao, Lauren Hong, Sarah Pertsemlidis, Mayumi Shaepers-Cheu, Tian~Zi Wang, Divya Srijay, Connor Monticello, Pranay Vure, Rishab Pulugurta, Kseniia Kholina, Shrey Goel, Matthew~P DeLisa, Ray Truant, Hector~C Aguilar, and Pranam Chatterjee.
\newblock {PepMLM}: Target sequence-conditioned generation of therapeutic peptide binders via span masked language modeling.
\newblock \emph{ArXiv}, page arXiv:2310.03842v3, August 2024.

\bibitem[Creamer et~al.(2018)Creamer, Mano, and Clark]{Creamer2018-pg}
Matthew~S Creamer, Omer Mano, and Damon~A Clark.
\newblock Visual control of walking speed in drosophila.
\newblock \emph{Neuron}, 100\penalty0 (6):\penalty0 1460--1473.e6, December 2018.

\bibitem[Creamer et~al.(2024)Creamer, Leifer, and Pillow]{Creamer2024-sh}
Matthew~S Creamer, Andrew~M Leifer, and Jonathan~W Pillow.
\newblock Bridging the gap between the connectome and whole-brain activity in\textit{C. elegans}.
\newblock \emph{bioRxiv}, page 2024.09.22.614271, September 2024.

\bibitem[Cunningham et~al.(2023)Cunningham, Ewart, Riggs, Huben, and Sharkey]{Cunningham2023-ap}
Hoagy Cunningham, Aidan Ewart, Logan Riggs, Robert Huben, and Lee Sharkey.
\newblock Sparse autoencoders find highly interpretable features in language models.
\newblock \emph{arXiv [cs.LG]}, September 2023.

\bibitem[Denil et~al.(2014)Denil, Demiraj, and de~Freitas]{Denil2014-wb}
Misha Denil, Alban Demiraj, and Nando de~Freitas.
\newblock Extraction of salient sentences from labelled documents.
\newblock \emph{arXiv [cs.CL]}, December 2014.

\bibitem[Devlin et~al.(2018)Devlin, Chang, Lee, and Toutanova]{Devlin2018-ym}
Jacob Devlin, Ming-Wei Chang, Kenton Lee, and Kristina Toutanova.
\newblock {BERT}: Pre-training of deep bidirectional transformers for language understanding.
\newblock \emph{arXiv [cs.CL]}, October 2018.

\bibitem[Dixen et~al.(2025)Dixen, Heinrich, and Burelli]{Dixen2025-fl}
Laurits Dixen, Stefan Heinrich, and Paolo Burelli.
\newblock Exploring deep learning models for {EEG} neural decoding.
\newblock \emph{arXiv [cs.LG]}, March 2025.

\bibitem[Dominey et~al.(2022)Dominey, Ellmore, and Ventre-Dominey]{Dominey2022-nz}
Peter~Ford Dominey, Timothy~M Ellmore, and Jocelyne Ventre-Dominey.
\newblock Effects of connectivity on narrative temporal processing in structured reservoir computing.
\newblock In \emph{2022 International Joint Conference on Neural Networks (IJCNN)}, pages 1--8, July 2022.

\bibitem[Fouke et~al.(2025)Fouke, He, Loring, and Naumann]{Fouke2025-xy}
Kaitlyn~E Fouke, Zichen He, Matthew~D Loring, and Eva~A Naumann.
\newblock Neural circuits underlying divergent visuomotor strategies of zebrafish and danionella cerebrum.
\newblock \emph{Curr. Biol.}, May 2025.

\bibitem[Gauthier et~al.(2021)Gauthier, Bollt, Griffith, and Barbosa]{Gauthier2021-yv}
Daniel~J Gauthier, Erik Bollt, Aaron Griffith, and Wendson A~S Barbosa.
\newblock Next generation reservoir computing.
\newblock \emph{Nat. Commun.}, 12\penalty0 (1):\penalty0 5564, September 2021.

\bibitem[Giovannucci et~al.(2019)Giovannucci, Friedrich, Gunn, Kalfon, Brown, Koay, Taxidis, Najafi, Gauthier, Zhou, Khakh, Tank, Chklovskii, and Pnevmatikakis]{Giovannucci2019-ix}
Andrea Giovannucci, Johannes Friedrich, Pat Gunn, Jérémie Kalfon, Brandon~L Brown, Sue~Ann Koay, Jiannis Taxidis, Farzaneh Najafi, Jeffrey~L Gauthier, Pengcheng Zhou, Baljit~S Khakh, David~W Tank, Dmitri~B Chklovskii, and Eftychios~A Pnevmatikakis.
\newblock {CaImAn} an open source tool for scalable calcium imaging data analysis.
\newblock \emph{Elife}, 8, January 2019.

\bibitem[Gudapati et~al.(2020)Gudapati, Singh, Clarkson-Townsend, Phillips, Douglass, Feola, and Allen]{Gudapati2020-jl}
Kaavya Gudapati, Anayesha Singh, Danielle Clarkson-Townsend, Stephen~Q Phillips, Amber Douglass, Andrew~J Feola, and Rachael~S Allen.
\newblock Behavioral assessment of visual function via optomotor response and cognitive function via {Y}-maze in diabetic rats.
\newblock \emph{J. Vis. Exp.}, \penalty0 (164):\penalty0 10.3791/61806, October 2020.

\bibitem[Guo et~al.(2024)Guo, Zhu, Yang, Xie, Dong, Zhang, Chen, Bi, Wu, Li, Luo, Xiong, and Liang]{Guo2024-dk}
Daya Guo, Qihao Zhu, Dejian Yang, Zhenda Xie, Kai Dong, Wentao Zhang, Guanting Chen, Xiao Bi, Y~Wu, Y~K Li, Fuli Luo, Yingfei Xiong, and Wenfeng Liang.
\newblock {DeepSeek}-coder: When the large language model meets programming -- the rise of code intelligence.
\newblock \emph{arXiv [cs.SE]}, January 2024.

\bibitem[Haesemeyer et~al.(2018)Haesemeyer, Robson, Li, Schier, and Engert]{Haesemeyer2018-fr}
Martin Haesemeyer, Drew~N Robson, Jennifer~M Li, Alexander~F Schier, and Florian Engert.
\newblock A brain-wide circuit model of heat-evoked swimming behavior in larval zebrafish.
\newblock \emph{Neuron}, 98\penalty0 (4):\penalty0 817--831.e6, May 2018.

\bibitem[Hou and Castanon(2023)]{Hou2023-ls}
Elizabeth~M Hou and Gregory Castanon.
\newblock Decoding layer saliency in language transformers.
\newblock \emph{arXiv [cs.CL]}, August 2023.

\bibitem[Huang et~al.(2020)Huang, Yan, Wang, Li, Yang, Li, Zuo, Zhang, and Chen]{Huang2020-tt}
Wei Huang, Hongmei Yan, Chong Wang, Jiyi Li, Xiaoqing Yang, Liang Li, Zhentao Zuo, Jiang Zhang, and Huafu Chen.
\newblock Long short-term memory-based neural decoding of object categories evoked by natural images.
\newblock \emph{Hum. Brain Mapp.}, 41\penalty0 (15):\penalty0 4442--4453, October 2020.

\bibitem[Kingma and Ba(2014)]{Kingma2014-zp}
Diederik~P Kingma and Jimmy Ba.
\newblock Adam: A method for stochastic optimization.
\newblock \emph{arXiv [cs.LG]}, December 2014.

\bibitem[Lappalainen et~al.(2024)Lappalainen, Tschopp, Prakhya, McGill, Nern, Shinomiya, Takemura, Gruntman, Macke, and Turaga]{Lappalainen2024-ip}
Janne~K Lappalainen, Fabian~D Tschopp, Sridhama Prakhya, Mason McGill, Aljoscha Nern, Kazunori Shinomiya, Shin-Ya Takemura, Eyal Gruntman, Jakob~H Macke, and Srinivas~C Turaga.
\newblock Connectome-constrained networks predict neural activity across the fly visual system.
\newblock \emph{Nature}, 634\penalty0 (8036):\penalty0 1132--1140, October 2024.

\bibitem[Liu et~al.(2023)Liu, Liu, Gao, Gao, Zhao, Li, Ding, and Wen]{Liu2023-yc}
Peiyu Liu, Zikang Liu, Ze-Feng Gao, Dawei Gao, Wayne~Xin Zhao, Yaliang Li, Bolin Ding, and Ji-Rong Wen.
\newblock Do emergent abilities exist in quantized large language models: An empirical study.
\newblock \emph{arXiv [cs.CL]}, July 2023.

\bibitem[Loshchilov and Hutter(2017)]{Loshchilov2017-hk}
Ilya Loshchilov and Frank Hutter.
\newblock Decoupled weight decay regularization.
\newblock \emph{arXiv [cs.LG]}, November 2017.

\bibitem[Lueckmann et~al.(2025)Lueckmann, Immer, Chen, Li, Petkova, Iyer, Hesselink, Dev, Ihrke, Park, Petruncio, Weigel, Korff, Engert, Lichtman, Ahrens, Januszewski, and Jain]{Lueckmann2025-ze}
Jan-Matthis Lueckmann, Alexander Immer, Alex Bo-Yuan Chen, Peter~H Li, Mariela~D Petkova, Nirmala~A Iyer, Luuk~Willem Hesselink, Aparna Dev, Gudrun Ihrke, Woohyun Park, Alyson Petruncio, Aubrey Weigel, Wyatt Korff, Florian Engert, Jeff~W Lichtman, Misha~B Ahrens, Michał Januszewski, and Viren Jain.
\newblock {ZAPBench}: A benchmark for whole-brain activity prediction in zebrafish.
\newblock \emph{arXiv [q-bio.NC]}, March 2025.

\bibitem[Mathis et~al.(2024)Mathis, Perez~Rotondo, Chang, Tolias, and Mathis]{Mathis2024-fm}
Mackenzie~Weygandt Mathis, Adriana Perez~Rotondo, Edward~F Chang, Andreas~S Tolias, and Alexander Mathis.
\newblock Decoding the brain: From neural representations to mechanistic models.
\newblock \emph{Cell}, 187\penalty0 (21):\penalty0 5814--5832, October 2024.

\bibitem[Morra et~al.(2025)Morra, Fouke, Naumann, and Daley]{Morra2025-pf}
Jacob Morra, Kaitlyn Fouke, Eva~A Naumann, and Mark Daley.
\newblock Connectomes inform function: from time-varying dynamics to animal behaviour.
\newblock \emph{Nat. Comput.}, pages 1--18, June 2025.

\bibitem[Musall et~al.(2019)Musall, Urai, Sussillo, and Churchland]{Musall2019-rv}
Simon Musall, Anne~E Urai, David Sussillo, and Anne~K Churchland.
\newblock Harnessing behavioral diversity to understand neural computations for cognition.
\newblock \emph{Curr. Opin. Neurobiol.}, 58:\penalty0 229--238, October 2019.

\bibitem[Naumann et~al.(2016)Naumann, Fitzgerald, Dunn, Rihel, Sompolinsky, and Engert]{Naumann2016-mx}
Eva~A Naumann, James~E Fitzgerald, Timothy~W Dunn, Jason Rihel, Haim Sompolinsky, and Florian Engert.
\newblock From whole-brain data to functional circuit models: The zebrafish optomotor response.
\newblock \emph{Cell}, 167\penalty0 (4):\penalty0 947--960.e20, November 2016.

\bibitem[Ng(2011)]{Ng2011Sparse}
Andrew~Y. Ng.
\newblock Sparse autoencoders.
\newblock CS294A Lecture Notes~72, Stanford University, 2011.
\newblock URL \url{https://web.stanford.edu/class/cs294a/sparseAutoencoder.pdf}.

\bibitem[Pachitariu et~al.(2016)Pachitariu, Stringer, Dipoppa, Schröder, Rossi, Dalgleish, Carandini, and Harris]{Pachitariu2016-vp}
Marius Pachitariu, Carsen Stringer, Mario Dipoppa, Sylvia Schröder, L~Federico Rossi, Henry Dalgleish, Matteo Carandini, and Kenneth~D Harris.
\newblock {Suite2p}: beyond 10,000 neurons with standard two-photon microscopy.
\newblock \emph{bioRxiv}, page 061507, June 2016.

\bibitem[Paszke et~al.(2019)Paszke, Gross, Massa, Lerer, Bradbury, Chanan, Killeen, Lin, Gimelshein, Antiga, Desmaison, Köpf, Yang, DeVito, Raison, Tejani, Chilamkurthy, Steiner, Fang, Bai, and Chintala]{Paszke2019-pq}
Adam Paszke, Sam Gross, Francisco Massa, Adam Lerer, James Bradbury, Gregory Chanan, Trevor Killeen, Zeming Lin, Natalia Gimelshein, Luca Antiga, Alban Desmaison, Andreas Köpf, Edward Yang, Zach DeVito, Martin Raison, Alykhan Tejani, Sasank Chilamkurthy, Benoit Steiner, Lu~Fang, Junjie Bai, and Soumith Chintala.
\newblock {PyTorch}: An imperative style, high-performance deep learning library.
\newblock \emph{arXiv [cs.LG]}, December 2019.

\bibitem[Radford et~al.(2019)Radford, Wu, Child, Luan, Amodei, and Sutskever]{Radford2019-kc}
Alec Radford, Jeffrey Wu, Rewon Child, David Luan, Dario Amodei, and Ilya Sutskever.
\newblock Language models are unsupervised multitask learners.
\newblock February 2019.

\bibitem[Rae et~al.(2021)Rae, Borgeaud, Cai, Millican, Hoffmann, Song, Aslanides, Henderson, Ring, Young, Rutherford, Hennigan, Menick, Cassirer, Powell, van~den Driessche, Hendricks, Rauh, Huang, Glaese, Welbl, Dathathri, Huang, Uesato, Mellor, Higgins, Creswell, McAleese, Wu, Elsen, Jayakumar, Buchatskaya, Budden, Sutherland, Simonyan, Paganini, Sifre, Martens, Li, Kuncoro, Nematzadeh, Gribovskaya, Donato, Lazaridou, Mensch, Lespiau, Tsimpoukelli, Grigorev, Fritz, Sottiaux, Pajarskas, Pohlen, Gong, Toyama, d'Autume, Li, Terzi, Mikulik, Babuschkin, Clark, Casas, Guy, Jones, Bradbury, Johnson, Hechtman, Weidinger, Gabriel, Isaac, Lockhart, Osindero, Rimell, Dyer, Vinyals, Ayoub, Stanway, Bennett, Hassabis, Kavukcuoglu, and Irving]{Rae2021-nq}
Jack~W Rae, Sebastian Borgeaud, Trevor Cai, Katie Millican, Jordan Hoffmann, Francis Song, John Aslanides, Sarah Henderson, Roman Ring, Susannah Young, Eliza Rutherford, Tom Hennigan, Jacob Menick, Albin Cassirer, Richard Powell, George van~den Driessche, Lisa~Anne Hendricks, Maribeth Rauh, Po-Sen Huang, Amelia Glaese, Johannes Welbl, Sumanth Dathathri, Saffron Huang, Jonathan Uesato, John Mellor, Irina Higgins, Antonia Creswell, Nat McAleese, Amy Wu, Erich Elsen, Siddhant Jayakumar, Elena Buchatskaya, David Budden, Esme Sutherland, Karen Simonyan, Michela Paganini, Laurent Sifre, Lena Martens, Xiang~Lorraine Li, Adhiguna Kuncoro, Aida Nematzadeh, Elena Gribovskaya, Domenic Donato, Angeliki Lazaridou, Arthur Mensch, Jean-Baptiste Lespiau, Maria Tsimpoukelli, Nikolai Grigorev, Doug Fritz, Thibault Sottiaux, Mantas Pajarskas, Toby Pohlen, Zhitao Gong, Daniel Toyama, Cyprien de~Masson d'Autume, Yujia Li, Tayfun Terzi, Vladimir Mikulik, Igor Babuschkin, Aidan Clark, Diego de~Las Casas, Aurelia Guy, Chris Jones,
  James Bradbury, Matthew Johnson, Blake Hechtman, Laura Weidinger, Iason Gabriel, William Isaac, Ed~Lockhart, Simon Osindero, Laura Rimell, Chris Dyer, Oriol Vinyals, Kareem Ayoub, Jeff Stanway, Lorrayne Bennett, Demis Hassabis, Koray Kavukcuoglu, and Geoffrey Irving.
\newblock Scaling language models: Methods, analysis \& insights from training gopher.
\newblock \emph{arXiv [cs.CL]}, December 2021.

\bibitem[Serrano and Smith(2019)]{Serrano2019-wk}
Sofia Serrano and Noah~A Smith.
\newblock Is attention interpretable?
\newblock \emph{arXiv [cs.CL]}, June 2019.

\bibitem[Shaw et~al.(2018)Shaw, Uszkoreit, and Vaswani]{Shaw2018-jo}
Peter Shaw, Jakob Uszkoreit, and Ashish Vaswani.
\newblock Self-attention with relative position representations.
\newblock \emph{arXiv [cs.CL]}, March 2018.

\bibitem[Shevchenko et~al.(2024)Shevchenko, Yeremeieva, and Laschowski]{Shevchenko2024-ta}
Olena Shevchenko, Sofiia Yeremeieva, and Brokoslaw Laschowski.
\newblock Comparative analysis of neural decoding algorithms for brain-machine interfaces.
\newblock \emph{bioRxiv}, page 2024.12.05.627080, December 2024.

\bibitem[Shi et~al.(2018)Shi, Yuan, Chang, Cho, Xie, Chen, and Luo]{Shi2018-hf}
Cong Shi, Xuedong Yuan, Karen Chang, Kin-Sang Cho, Xinmin~Simon Xie, Dong~Feng Chen, and Gang Luo.
\newblock Optimization of optomotor response-based visual function assessment in mice.
\newblock \emph{Sci. Rep.}, 8\penalty0 (1):\penalty0 9708, June 2018.

\bibitem[Simonyan et~al.(2013)Simonyan, Vedaldi, and Zisserman]{Simonyan2013-lu}
Karen Simonyan, Andrea Vedaldi, and Andrew Zisserman.
\newblock Deep inside convolutional networks: Visualising image classification models and saliency maps.
\newblock \emph{arXiv [cs.CV]}, December 2013.

\bibitem[Tayebi et~al.(2023)Tayebi, Azadnajafabad, Maroufi, Pour-Rashidi, Khorasanizadeh, Faramarzi, and Slavin]{Tayebi2023-ed}
Hossein Tayebi, Sina Azadnajafabad, Seyed~Farzad Maroufi, Ahmad Pour-Rashidi, Mirhojjat Khorasanizadeh, Sina Faramarzi, and Konstantin~V Slavin.
\newblock Applications of brain-computer interfaces in neurodegenerative diseases.
\newblock \emph{Neurosurg. Rev.}, 46\penalty0 (1):\penalty0 131, May 2023.

\bibitem[Tschopp et~al.(2018)Tschopp, Reiser, and Turaga]{Tschopp2018-iu}
Fabian~David Tschopp, Michael~B Reiser, and Srinivas~C Turaga.
\newblock A connectome based hexagonal lattice convolutional network model of the drosophila visual system.
\newblock \emph{arXiv [q-bio.NC]}, June 2018.

\bibitem[Urai et~al.(2022)Urai, Doiron, Leifer, and Churchland]{Urai2022-cf}
Anne~E Urai, Brent Doiron, Andrew~M Leifer, and Anne~K Churchland.
\newblock Large-scale neural recordings call for new insights to link brain and behavior.
\newblock \emph{Nat. Neurosci.}, 25\penalty0 (1):\penalty0 11--19, January 2022.

\bibitem[Vaswani et~al.(2017)Vaswani, Shazeer, Parmar, Uszkoreit, Jones, Gomez, Kaiser, and Polosukhin]{Vaswani2017-nt}
Ashish Vaswani, Noam Shazeer, Niki Parmar, Jakob Uszkoreit, Llion Jones, Aidan~N Gomez, Lukasz Kaiser, and Illia Polosukhin.
\newblock Attention is all you need.
\newblock \emph{arXiv [cs.CL]}, June 2017.

\bibitem[Vyas et~al.(2020)Vyas, Golub, Sussillo, and Shenoy]{Vyas2020-wr}
Saurabh Vyas, Matthew~D Golub, David Sussillo, and Krishna~V Shenoy.
\newblock Computation through neural population dynamics.
\newblock \emph{Annu. Rev. Neurosci.}, 43\penalty0 (1):\penalty0 249--275, July 2020.

\bibitem[Wang et~al.(2023)Wang, Subramaniam, Yaari, Kreiman, Katz, Cases, and Barbu]{Wang2023-ge}
Christopher Wang, Vighnesh Subramaniam, Adam~Uri Yaari, Gabriel Kreiman, Boris Katz, Ignacio Cases, and Andrei Barbu.
\newblock {BrainBERT}: Self-supervised representation learning for intracranial recordings.
\newblock \emph{arXiv [cs.LG]}, February 2023.

\bibitem[Wang et~al.(2022)Wang, Li, Wu, Hovy, and Sun]{Wang2022-vj}
Haifeng Wang, Jiwei Li, Hua Wu, Eduard Hovy, and Yu~Sun.
\newblock Pre-trained language models and their applications.
\newblock \emph{Engineering (Beijing)}, 25:\penalty0 51--65, September 2022.

\bibitem[Wang and Jiang(2015)]{Wang2015-uu}
Shuohang Wang and Jing Jiang.
\newblock Learning natural language inference with {LSTM}.
\newblock \emph{arXiv [cs.CL]}, December 2015.

\bibitem[Wei et~al.(2022)Wei, Tay, Bommasani, Raffel, Zoph, Borgeaud, Yogatama, Bosma, Zhou, Metzler, Chi, Hashimoto, Vinyals, Liang, Dean, and Fedus]{Wei2022-kl}
Jason Wei, Yi~Tay, Rishi Bommasani, Colin Raffel, Barret Zoph, Sebastian Borgeaud, Dani Yogatama, Maarten Bosma, Denny Zhou, Donald Metzler, Ed~H Chi, Tatsunori Hashimoto, Oriol Vinyals, Percy Liang, Jeff Dean, and William Fedus.
\newblock Emergent abilities of large language models.
\newblock \emph{arXiv [cs.CL]}, June 2022.

\bibitem[Xiong et~al.(2021)Xiong, Zeng, Chakraborty, Tan, Fung, Li, and Singh]{Xiong2021-yu}
Yunyang Xiong, Zhanpeng Zeng, Rudrasis Chakraborty, Mingxing Tan, Glenn Fung, Yin Li, and Vikas Singh.
\newblock Nystromformer: A nystrom-based algorithm for approximating self-attention.
\newblock \emph{arXiv [cs.CL]}, February 2021.

\bibitem[Ye and Pandarinath(2021)]{Ye2021-ry}
Joel Ye and Chethan Pandarinath.
\newblock Representation learning for neural population activity with neural data transformers.
\newblock \emph{bioRxiv}, January 2021.

\bibitem[Ye et~al.(2023)Ye, Collinger, Wehbe, and Gaunt]{Ye2023-ku}
Joel Ye, Jennifer~L Collinger, Leila Wehbe, and Robert Gaunt.
\newblock Neural data transformer 2: Multi-context pretraining for neural spiking activity.
\newblock \emph{bioRxivorg}, page 2023.09.18.558113, September 2023.

\bibitem[Zhang et~al.(2025)Zhang, Wang, Feng, Tan, Liu, and Tsvetkov]{Zhang2025-yq}
Yizhuo Zhang, Heng Wang, Shangbin Feng, Zhaoxuan Tan, Xinyun Liu, and Yulia Tsvetkov.
\newblock Generalizable {LLM} learning of graph synthetic data with reinforcement learning.
\newblock \emph{arXiv [cs.LG]}, June 2025.

\bibitem[Zhang et~al.(2024)Zhang, Wang, Jimenez-Beneto, Wang, Azabou, Richards, Winter, {International Brain Laboratory}, Dyer, Paninski, and Hurwitz]{Zhang2024-je}
Yizi Zhang, Yanchen Wang, Donato Jimenez-Beneto, Zixuan Wang, Mehdi Azabou, Blake Richards, Olivier Winter, {International Brain Laboratory}, Eva Dyer, Liam Paninski, and Cole Hurwitz.
\newblock Towards a ``universal translator'' for neural dynamics at single-cell, single-spike resolution.
\newblock \emph{arXiv [q-bio.NC]}, July 2024.

\bibitem[Zhang et~al.(2023)Zhang, Huang, Hu, Zhao, Wang, Liu, Zhang, Joe~Qin, and Zhao]{Zhang2023-uf}
Zijian Zhang, Ze~Huang, Zhiwei Hu, Xiangyu Zhao, Wanyu Wang, Zitao Liu, Junbo Zhang, S~Joe~Qin, and Hongwei Zhao.
\newblock {MLPST}: {MLP} is all you need for spatio-temporal prediction.
\newblock \emph{arXiv [cs.LG]}, September 2023.

\bibitem[Štih et~al.(2019)Štih, Petrucco, Kist, and Portugues]{Stih2019-uv}
Vilim Štih, Luigi Petrucco, Andreas~M Kist, and Ruben Portugues.
\newblock Stytra: An open-source, integrated system for stimulation, tracking and closed-loop behavioral experiments.
\newblock \emph{PLoS Comput. Biol.}, 15\penalty0 (4):\penalty0 e1006699, April 2019.

\end{thebibliography}

%%%%%%%%%%%%%%%%%%%%%%%%%%%%%%%%%%%%%%%%%%%%%%%%%%%%%%%%%%%%

\appendix
\newpage

\section{Technical Appendices and Supplementary Material}\label{Sec:Appendix}
\subsection{Data acquisition}
Each fish was presented with visual motion stimuli from directly underneath, as shown in Fig. \ref{fig:data_acqu}. This fish was behaving, i.e. engaged in the optomotor response (OMR), as described in \cite{Naumann2016-mx, Shi2018-hf, Gudapati2020-jl}. The suite of OMR stimuli include whole field moving gratings presented at angles of 0°, 45°, 90°, 135°, 180°, 225°, 270°, or 315° relative to the fish’s orientation, with 0° denoting tail to head motion, as well as leftward and rightward motion separately to each eye. Each stimulus lasted 40 s, beginning with a 30 s stationary period during which the orientation of the grating changed, followed by a motion for 10 s, long enough for neural activity to reach characteristic peaks and return to baseline and allow for behavioral observation associated with each stimulus.

\subsection{Computing setup}
All experiments were run using a Nvidia RTX A6000 GPU, housed remotely and accessed as a cloud resource via command line interface. Model access makes use of a SLURM (Simple Linux Utility for Resource Management) workload manager. All code and data are available at \href{https://github.com/Naumann-Lab/nlp4neuro.git}{https://github.com/Naumann-Lab/nlp4neuro.git} and \href{https://doi.org/10.7910/DVN/I8LULX}{https://doi.org/10.7910/DVN/I8LULX}.

\subsection{Reservoir computing}
The RC uses a random, fixed reservoir layer $\mathbf{W} \in \mathbb{R}^{n \times n}$, which drives an input time series $u(t)$, scaled by an input-to-reservoir matrix $\mathbf{W}_{in}$, towards a set of $n$ activity vectors $x(t)$, where for some leaking rate $\gamma$:

\begin{equation}\label{eqn: rc1}
\dot x(t) = -\gamma x(t) + tanh(\mathbf{W}x(t) + \mathbf{W}_{in}u(t)).
\end{equation}

A linear combination of these activity vectors are used to minimize the Mean-squared error (MSE) between a known ground truth signal $y(t)$ and the predicted output signal $\hat{y}(t)$, in one-shot. These learned weights are contained in $\mathbf{W}_{out}$, which is computed using:

\begin{equation}\label{eqn: rc2}
\mathbf{W}_{out} = Y X^\top (X X^\top + \alpha I)^{-1}
\end{equation}

where $X$ is the matrix of all activity vectors $x(t)$, $Y$ is the matrix of all ground truth outputs $y(t)$, and $\alpha$ is a regularization parameter. We adapt this model such that $u(t) = s_1(t)$ for a sequence $s_1$ of neural data and such that $y(t) = s_2(t)$ for a concurrent sequence of behavioral data $s_2(t)$.

\subsection{Long Short-Term Memory networks (LSTMs)}
The LSTM comprises three gates, the input gate $i(t)$, output gate $o(t)$, and forget gate $f(t)$, and two state vectors, a cell state $c(t)$ and hidden state $h(t)$, which update per each time step $t$, given an input time series $u(t)$, according to the following equations:

\noindent
\begin{align}
\hspace*{-1cm}\textbf{Gate Equation:}\quad
g(t) &= \sigma\Big(W_g\, u(t) + U_g\, h(t-1) + b_g\Big), 
\label{eq:input_gate} \\[1ex]
\hspace*{-1cm}\textbf{Cell State Update:}\quad
c(t) &= f(t) * c(t-1) + i(t) * \tilde{c}(t), 
\label{eq:cell_update} \\[1ex]
\hspace*{-1cm}\textbf{Hidden State:}\quad
h(t) &= o(t) * \tanh\big(c(t)\big).
\label{eq:hidden_state}
\end{align}
where \( W_g\) and \( U_g \) are weight matrices, $b_g$ is a bias term, \( \sigma(\cdot) \) is the sigmoid function, and 
\begin{align}
\tilde{c}(t) &= \tanh\Big(W_c\, u(t) + U_c\, h(t-1) + b_c\Big).
\label{eq:candidate_state}
\end{align}

\subsection{BERT/transformer implementation}
\textbf{1. Multi-Headed Self-Attention:}

For all tokens $u(t) \in \mathbb{R}^{n,m}$, we construct a key matrix $K$, query matrix $Q$, and value matrix $V$, where 
\begin{equation}\label{eq:qkv}
Q(t) = u(t)W^Q,\quad K(t)=u(t)W^K,\quad V(t)=u(t)W^V.    
\end{equation}

To determine which tokens in the sequence should be attended to, based on the context of a given token within a context window, BERT computes attention weights using: 

\begin{equation}\label{eq:attn}
\text{Attention} = \mathrm{softmax}_j\!\Big(\frac{Q(t)K(j)^\top}{\sqrt{d_k}}\Big),
\end{equation}

before taking a linear combination of all attention weights applied to value matrix $V$.

\begin{equation}\label{eq:attnfinal}
z(t) = \sum_{j=1}^{T} a_{tj}\, V(j).    
\end{equation}

\textbf{2. Layer Normalization}
After applying multi-headed attention, and generating $z(t)$, we simply update the existing token $u(t)$ with $z(t)$ and apply layer normalization.

\begin{equation}\label{eq:layernorm}
u'(t) = \mathrm{LayerNorm}\Big(u(t) + z(t)\Big).
\end{equation}

\textbf{3. Feed-Forward Network}
Finally, we run the transformed token $u'(t)$ through a multi-layer perceptron (MLP) which reduces to the dimensionality of the input token. In BERT, the gaussian error linear unit, or GELU activation function is used, as described in \cite{Devlin2018-ym}.

\begin{equation}\label{eq:attnmlp}
\tilde{u}(t) = \mathrm{GELU}\Big(u'(t)W_1 + b_1\Big)W_2 + b_2.
\end{equation}

Finally, an additional layer normalization is applied to the token before it is passed onto another transformer block.

\subsection{Mixture of Experts (MoE)}

MoE models adopt a learnable routing matrix $W_r$, with routing $r(t)$ at time $t$ denoted by:

\begin{equation}\label{eq:routing}
r(t) = \mathrm{softmax}\Big(W_{r}\, h(t) + b_{r}\Big).   
\end{equation}

As mentioned previously, the MoE model recruits the top-$k$ experts ($e_i$) with indices $i$ such that

\begin{equation}\label{eq:moe}
z(t) = \frac{1}{k} \sum_{i=1}^{k} \mathrm{GELU}\big(\big( h(t)W_{e_i}^{(1)}+b_{e_i}^{(1))} \big)W_{e_i}^{(2)} + b_{e_i}^{(1)}\big).
\end{equation}

In our DeepSeek-c7b implementation, we use 4-bit (Exp. 3) and 8-bit (Exp. 1, 2 and 4) quantized versions of the model, to reduce memory requirements during fine-tuning and from-scratch training.

\subsection{Evaluation}
We used root-mean squared error for all model evaluations, for assessing ground truth $u(t)$ and predicted data $\hat{u}(t)$, given $T$ total time steps.

\begin{equation}\label{eqn:rmse}
    \text{RMSE} = \sqrt{\frac{1}{T} \sum_{t=1}^{T} \left( u(t) - \hat{u}(t) \right)^2}
\end{equation}

We also consider Pearson's coefficient of correlation, $r$. Pearson's product-moment correlation coefficient, $r$, between model tail sum predictions $\theta_{\text{sum},p}$ and ground truth tail sum data $\theta_{\text{sum},p}$, where $\theta_{\text{sum}}$ adds each tail component measurement $\theta_i$ for $i \in [0,6)$:

\begin{equation}
\hspace{-.7cm}
r = \frac{\sum_{t=1}^{T} \Bigl(\theta_{\text{sum},p}(t) - \overline{\theta}_{\text{sum},p}\Bigr)
\Bigl(\theta_{\text{sum},g}(t) - \overline{\theta}_{\text{sum},g}\Bigr)}
{\sqrt{\sum_{t=1}^{T} \Bigl(\theta_{\text{sum},p}(t) - \overline{\theta}_{\text{sum},p}\Bigr)^2} \,
\sqrt{\sum_{t=1}^{T} \Bigl(\theta_{\text{sum},g}(t) - \overline{\theta}_{\text{sum},g}\Bigr)^2}},
\label{eq:pearson_thetasum_time}
\end{equation}

where $T$ is the size of the test set, and $t$ is the frame step.

\subsection{Embedding strategies}

We have adopted the classic positional encoding strategy from \cite{Vaswani2017-nt}. That is, for a neural data token $n_t$, here of dimension $d_{\text{model}}$=15,081, we add a positional encoding vector $PE$, where for token position $\mathbf{n}_t$ and row index $i$:

\begin{equation}
\text{PE}(p, 2i) = \sin\!\left(\frac{p}{10000^{\frac{2i}{d_{\text{model}}}}}\right), \quad i = 0, 1, 2, \dots
\label{eq:pos_enc_even}
\end{equation}

\begin{equation}
\text{PE}(p, 2i+1) = \cos\!\left(\frac{p}{10000^{\frac{2i}{d_{\text{model}}}}}\right), \quad i = 0, 1, 2, \dots
\label{eq:pos_enc_odd}
\end{equation}

The updated token $\mathbf{n}'_t$, is written simply as:
\begin{equation}
\mathbf{n}'_t = \mathbf{n}_t + \text{PE}(p)
\label{eq:add_pos_enc}
\end{equation}

The above is extended in relative positional encoding to account for all pairwise positions.

Regarding dimensionality reduction techniques used to encode neural data tokens, first, we considered a network-based embedding: \textit{Laplacian eigenmaps}, as described in \cite{Belkin2003-ry}, to construct a computational graph from an $n \times m$ input dataset $\mathcal{D}$, where each sample is a node in the graph. A correlation or similarity metric is used to assign weighted edges between nodes. The graph Laplacian $L$ of the network is computed (see Eq. \ref{eqn:glap}), and the smallest $k$ non-zero eigenvalues of $L$ are used to project the data onto the corresponding eigenvectors. 

\begin{equation}
L_{ij} =
\begin{cases} 
\text{deg}(i) & \text{if } i = j, \\
-1 & \text{if } i \neq j \text{ and nodes } i \text{ and } j \text{ are adjacent}, \\
0 & \text{otherwise}.
\end{cases}
\label{eqn:glap}
\end{equation}

Second, we considered a neural network-based embedding: sparse autoencoders. An autoencoder learns a latent representation of an input space (encoder), such that it can be reconstructed (decoder) into its original form. The decoder seeks to minimize the difference between the original input and its reconstruction. In \textit{sparse autoencoders}, a sparsity constraint (e.g. via L1 regularization) is applied to the encoder block of the network, which pushes the activity of some encoder neurons to 0, cutting their contribution to the learned input reconstruction, as noted in \cite{Cunningham2023-ap}.

\subsection{Salience mapping}
For $M$ windows containing frame $t$, each with window label $m$ and length $s$, the moment-to-moment salience for neuron $i$ at frame $t$, $S_{i,t}$, and global salience across all frames $T$, $S_i$, were computed as:

\begin{equation}\label{eqn:saliency}
  S_{i,t} = \frac{1}{M}\sum_{m}s^{(m)}_{i,t}, \qquad
  S_i = \frac{1}{T}\sum_{t=1}^{T} S_{i,t}.
\end{equation}

This approach is conceptually similar to classical gradient-based salience mapping strategies \cite{Simonyan2013-lu}, however with averaging across timesteps and windows to generate a per-neuron score. 

\subsection{Pre-trained versus un-trained model comparisons}

Table \ref{tab:pre_vs_un} provides the RMSE scores and Pearson's $r$-values corresponding to Fig. \ref{fig:pretrained_vs_untrained}.
\begin{table}[!htbp]
\centering
\caption{Pre-trained vs.\ un-trained models for sequence lengths $s=5$ and $s=20$. RMSE $\pm$~SEM and Pearson’s $r \pm$~SEM are reported. $^{\dagger}$ indicates a Wilcoxon signed‐rank $p<0.05$ between variants.}
\label{tab:pre_vs_un}
\begin{tabularx}{\textwidth}{@{} l l *{4}{>{\centering\arraybackslash}X} @{}}
\toprule
\textbf{Model} & \textbf{Metric} 
  & \multicolumn{2}{c}{$s=5$} 
  & \multicolumn{2}{c}{$s=20$} \\
\cmidrule(lr){3-4}\cmidrule(lr){5-6}
 &  & Pre-trained & Un-trained & Pre-trained & Un-trained \\
\midrule
BERT-bu       & RMSE & 0.060 $\pm$ 0.001 & 0.062 $\pm$ 0.002 & 0.055 $\pm$ 0.002$^{\dagger}$ & 0.063 $\pm$ 0.002 \\
           & $r$  & 0.05  $\pm$ 0.01  & 0.00  $\pm$ 0.01  & 0.06  $\pm$ 0.01  & 0.01  $\pm$ 0.01 \\
\addlinespace[0.6ex]
GPT-2      & RMSE & 0.060 $\pm$ 0.002$^{\dagger}$ & 0.060 $\pm$ 0.002 & 0.058 $\pm$ 0.002 & 0.052 $\pm$ 0.001 \\
           & $r$  & 0.07  $\pm$ 0.01  & 0.17  $\pm$ 0.02  & 0.04  $\pm$ 0.01  & 0.17  $\pm$ 0.02 \\
\addlinespace[0.6ex]
DeepSeek-c7b   & RMSE & 0.054 $\pm$ 0.003 & 0.060 $\pm$ 0.002 & 0.052 $\pm$ 0.001$^{\dagger}$ & 0.055 $\pm$ 0.002 \\
           & $r$  & 0.10  $\pm$ 0.01  & 0.11  $\pm$ 0.01  & 0.09  $\pm$ 0.01  & 0.12  $\pm$ 0.01 \\
\bottomrule
\end{tabularx}
\end{table}

\subsection{Performance for various embedding strategies}

Table \ref{tab:embeddings_results} illustrates the performance (RMSE) and standard error across LLMs and embedding strategies. 

\begin{table}[!h]
\centering
\footnotesize
\setlength\tabcolsep{4pt}
\caption{Experiment 3: Test performance for all embedding strategies. A \textbf{bolded} value marks the lowest RMSE in each column.}\label{tab:embeddings_results}
\begin{tabularx}{\textwidth}{@{} l l *{5}{>{\centering\arraybackslash}X} @{}}
\toprule
\textbf{Model} &  & \textbf{Vanilla} & \textbf{Positional} & \textbf{RelPos} & \textbf{Sparse} & \textbf{Spectral} \\
\midrule
BERT-bu   & RMSE & 0.120 ± 0.009 & 0.113 ± 0.007 & 0.112 ± 0.004 & 0.139 ± 0.006 & 0.109 ± 0.006 \\
\addlinespace[0.6ex]
GPT-2  & RMSE & 0.162 ± 0.024 & 0.146 ± 0.014 & 0.090 ± 0.008 & 0.185 ± 0.015 & 0.148 ± 0.021 \\
\addlinespace[0.6ex]
DeepSeek-c7b  & RMSE & \textbf{0.047 ± 0.003} & \textbf{0.046 ± 0.002}
                           & \textbf{0.045 ± 0.000} & \textbf{0.049 ± 0.003}
                           & \textbf{0.043 ± 0.002} \\
\addlinespace[0.6ex]
\bottomrule
\end{tabularx}
\end{table}

\subsection{Optomotor Stimulus presentation}

Optionally available for use in the dataset are the stimuli associated with each experiment. We provide these in \href{https://doi.org/10.7910/DVN/EFP1IL}{https://doi.org/10.7910/DVN/EFP1IL}.

\subsection{Principal Component Analysis}
We ran principal component analysis on high activity neurons, and on variants of DeepSeek-c7b which replace the $k$ most salient neurons with randomly sampled neurons. We find that the highest activity neurons affect embedding representations in PC-space.

\begin{figure}[!h]
    \centering
    \includegraphics[width=1\textwidth]{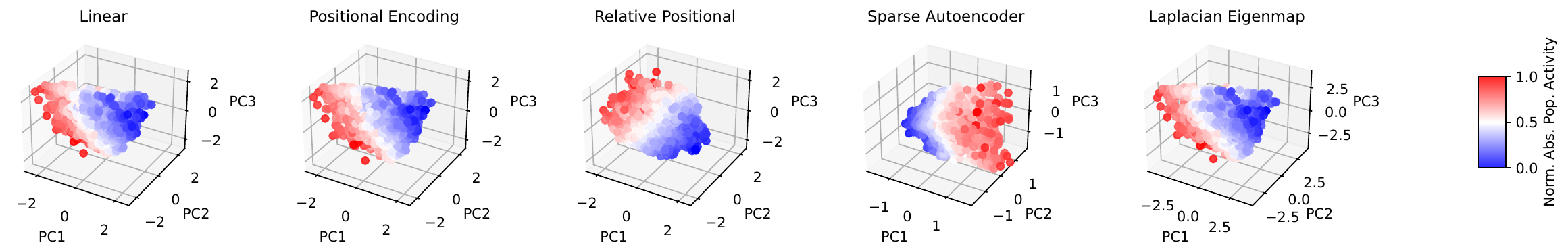}
    \caption{PCA applied to the full neural data token, with each point in PC-space corresponding to a frame for fish 1 neural data. Data is colored by absolute normalized activity for that time step.}
    \label{fig:enter-label}
\end{figure}
\begin{figure}[!h]
    \centering
    \includegraphics[width=1\textwidth]{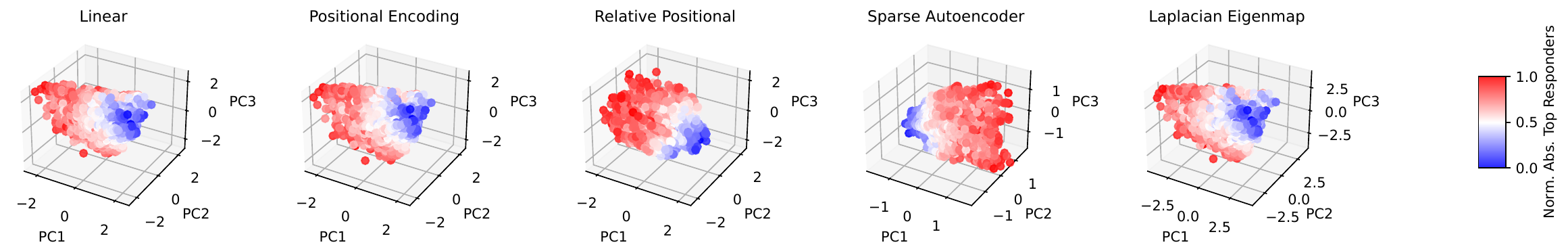}
    \caption{PCA applied to all neurons, with each point indicating a single frame of data for fish 1. When we ablate the top 50 highest activity neurons, certain representations (e.g. sparse autoencoder) appear less separable.}
    \label{fig:enter-label}
\end{figure}
\end{document}